\documentclass{aastex}

\begin{document}

\title{New Spectral Evidence of an Unaccounted Component of the Near-infrared \\
Extragalactic Background Light from the {\it CIBER}}

\author{S. Matsuura\altaffilmark{1,2}, T. Arai\altaffilmark{3,2}, J. Bock\altaffilmark{4,5}, A. Cooray\altaffilmark{6,4}, P. Korngut\altaffilmark{4,5}, M.G. Kim\altaffilmark{7,8}, H.M. Lee\altaffilmark{7}, D.H. Lee\altaffilmark{8,12}, L. Levenson\altaffilmark{4}, T. Matsumoto\altaffilmark{2,8}, Y. Onishi\altaffilmark{9,2}, M. Shirahata\altaffilmark{3,2}, K. Tsumura\altaffilmark{10}, T. Wada\altaffilmark{2} and M. Zemcov\altaffilmark{11,5}}

\email{matsuura.shuji@kwansei.ac.jp}

\altaffiltext{1}{School of Science and Technology, Kwansei Gakuin University, Sanda, Hyogo 669-1337, Japan}
\altaffiltext{2}{Department of Space Astronomy and Astrophysics, the Institute of Space and Astronautical Science, Japan Aerospace Exploration Agency, Sagamihara, Kanagawa 252-5210, Japan}
\altaffiltext{3}{Genesia Corporation, Mitaka, Tokyo 181-0013, Japan}
\altaffiltext{4}{Department of Physics, California Institute of Technology, CA 91125, USA}
\altaffiltext{5}{Jet Propulsion Laboratory, California Institute of Technology, CA 91109, USA}
\altaffiltext{6}{Center for Cosmology, University of California, Irvine, Irvine, CA 92697, USA}
\altaffiltext{7}{Department of Physics and Astronomy, Seoul National University, Seoul 151-742, Korea}
\altaffiltext{8}{Korea Astronomy and Space Science Institute, Daejeon 305-348, Korea}
\altaffiltext{9}{Graduate School of Science, Tokyo Institute of Technology, Tokyo 152-8550, Japan}
\altaffiltext{10}{Frontier Research Institute for Interdisciplinary Science, Tohoku University, Sendai, 980-8578, Japan}
\altaffiltext{11}{Center for Detectors, School of Physics and Astronomy, Rochester Institute of Technology, Rochester, NY 14623, USA}
\altaffiltext{12}{University of Science and Technology, Daejeon 34143, Korea}

\begin{abstract}

The Extragalactic Background Light (EBL) captures the total integrated emission from stars and galaxies throughout the cosmic history.  The amplitude of the near-infrared EBL from space absolute photometry observations has been controversial and depends strongly on the modeling and subtraction of the Zodiacal light foreground.  We report the first measurement of the diffuse background spectrum at 0.8--1.7 $\mu$m from the CIBER experiment.  The observations were obtained with an absolute spectrometer over two flights in multiple sky fields to enable the subtraction of Zodiacal light, stars, terrestrial emission, and diffuse Galactic light.  After subtracting foregrounds and accounting for systematic errors, we find the nominal EBL brightness, assuming the Kelsall Zodiacal light model, is 42.7$^{+11.9}_{-10.6}$ nW m$^{-2}$sr$^{-1}$ at 1.4 $\mu$m.  We also analyzed the data using the Wright Zodiacal light model, which results in a worse statistical fit to the data and an unphysical EBL, falling below the known background light from galaxies at $\lambda<$1.3 $\mu$m.  Using a model-independent analysis based on the minimum EBL brightness, we find an EBL brightness of 28.7$^{+5.1}_{-3.3}$ nWm$^{-2}$sr$^{-1}$ at 1.4 $\mu$m.  While the derived EBL amplitude strongly depends on the Zodiacal light model, we find that we cannot fit the spectral data to Zodiacal light, Galactic emission, and EBL from solely integrated galactic light from galaxy counts.  The results require a new diffuse component, such as an additional foreground or an excess EBL with a redder spectrum than that of Zodiacal light.

\end{abstract}

\keywords{cosmology: observations --- dark ages, reionization, first stars --- diffuse radiation --- infrared: diffuse background --- infrared: general --- zodiacal dust}

\section{Introduction}
The extragalactic background light (EBL) is the integrated intensity of all photons emitted and absorbed along a line of sight throughout the cosmic history.  While the early objects are too faint to detect individually, the EBL as the integrated light of such objects is potentially a powerful probe to search the spectral signatures.  As a result, the EBL is an important observable for understanding galaxy formation and evolution \citep{2001ARA&A..39..249H,2000MNRAS.312L...9M}.  Of particular interest, redshifted ultraviolet radiation from primordial galaxies and black holes during cosmic reionization has been proposed as a measurable component of the near-infrared EBL.  Such sources will contain a distinctive spectral Lyman break due to absorption by neutral hydrogen in the intergalactic medium and possibly a redshifted Lyman-alpha emission feature, leading to a spectral discontinuity at $\sim$1 $\mu$m for redshifts $z >$ 7 \citep{2002MNRAS.336.1082S,2003MNRAS.339..973S,2006ApJ...646..703F,2013MNRAS.431..383Y}.

Measurements of the EBL in the range 1.25 to 5 $\mu$m with {\it Cosmic Background Explorer} ({\it COBE}) \citep{1998ApJ...508...25H,2000ApJ...545...43W,2001ApJ...555..563C,2007ApJ...666...34L,2015ApJ...811...77S,2016ApJ...821L..11S}, {\it Infrared Telescope in Space} ({\it IRTS}) \citep{2005ApJ...626...31M,2015ApJ...807...57M} and {\it AKARI} \citep{2013PASJ...65..121T} have been difficult to reconcile with the integrated galaxy light (IGL) predicted from deep galaxy counts and contemporary models of galaxy evolution, exhibiting significantly larger surface brightnesses than predicted.  Interpretation of this EBL excess has remained controversial, with some identifying it with reionizing sources and others with local foregrounds.

A significant challenge to the measurement of the EBL near the earth is to accurately subtract the foreground zodiacal light (ZL), which is sunlight scattered by interplanetary dust in the plane of the planets.  The ZL has been modeled on the basis of the {\it COBE} all-sky observations \citep{1998ApJ...508...44K}, but since the ZL foreground dominates the EBL by factors $\gtrsim 20$ even a few percent uncertainty in the modeled ZL brightness changes the residual EBL level by large factors.  Over the 1.4--4 $\mu$m range reported by the IRTS \citep{2015ApJ...807...57M}, the EBL spectrum is similar to the ZL spectrum \citep{2005ApJ...635..784D}.  A promising approach to search for a reddening in the EBL spectrum, which would break the spectral degeneracy with ZL, is to extend the observation wavelengths to shorter than 1.25 $\mu$m.

In this paper, we report the first spectral measurement of the EBL in the unexplored wavelength range from 0.8 to 1.7 $\mu$m by a sounding rocket experiment, {\it Cosmic Infrared Background Experiment} ({\it CIBER}) \citep{2013ApJS..207...31Z,2013ApJS..207...32B,2013ApJS..207...33T,2013ApJS..207...34K}.

\section{{\it CIBER} experiment}
\subsection{The LRS instrument}
One of the science instruments of {\it CIBER}, the Low Resolution Spectrometer (LRS) is designed to measure the absolute spectrum of the near-infrared sky at wavelengths 0.8--2 $\mu$m.  A full description of the LRS is given in Tsumura et al. 2013a. The LRS consists of a 5-cm refractive telescope with five spectral slits providing 2.8-arcminute$\times$5.5-degree fields on the sky. A prism disperses the incident light perpendicular to the slits and the spectra are imaged with a 256$\times$256 HgCdTe detector array. Depending on the wavelength, a spectral resolving power of $\lambda$/$\delta\lambda$=15--30 is achieved.

\subsection{{\it CIBER} flights}\label{flights}
The {\it CIBER} experiment has flown four times.  The initial experiment flew on 10:45 UTC February 25, 2009 at White Sands Missile Range (WSMR), New Mexico, USA, on a two-stage Terrier-Black Brant IX rocket \footnote{ National Aerospace Administration (NASA) Sounding Rockets Program Office, The NASA Sounding Rocket Program Handbook (Wallops Island, VA: NASA)}.  Scattered thermal emission from the rocket skin \citep{2013ApJS..207...31Z} affected the LRS data, so these are not used in this EBL analysis, though we did identify a new spectral feature in the ZL from the data taken during this flight \citep{2010ApJ...719..394T}.

Following the first experiment, we refurbished the instrument to improve the baffling performance to scattered thermal emission.  The second experiment launched at 4:50 UTC July 11, 2010 from WSMR on a Terrier-Black Brant IX rocket.   As a result of the instrument improvements and performance of the vehicle, this flight generated the highest quality data of all the {\it CIBER} flights.

For the next flight, the experiment was modified to measure polarization of the ZL. The third experiment flew on 9:00 UTC March 22, 2012 from WSMR on a Terrier-Black Brant IX rocket.   Though we successfully measured the near-IR ZL polarization with these data, the photometric accuracy was not as good as that of the second flight due to strong airglow contamination.  We exclude this data set for this EBL analysis, though we plan to present ZL polarization results in a future paper.

For the fourth and final experiment, we removed the polarizer and modified the LRS to match the second flight configuration.  The fourth flight launched 3:05 UTC June 6, 2013 from Wallops Flight Facility (WFF), Virginia, USA, on a four-stage Black Brant XII rocket, which could carry the payload to much higher altitudes than the two-stage rocket used in the previous three flights.  In this experiment we obtained longer observation times on all science fields, and could evaluate residual airglow emission from the upper atmosphere over long time scales.

\subsection{Observations} \label{field}
Table~\ref{tbl-1} gives a summary of the science fields at varying equatorial, ecliptic and Galactic coordinates, and solar elongations in the three {\it CIBER} flights.  The coordinates were determined from the positions of {\it 2MASS} (Two Micron All Sky Survey) stars matched to the positions of stars observed with the LRS \citep{2017AJ....153...84K}.  

\begin{deluxetable}{lrrrrrrrrr}
\tabletypesize{\scriptsize}
\tablecaption{{\it CIBER/LRS} observed sky coordinates\label{tbl-1}}
\tablewidth{0pt}
\tablehead{
&&& J2000 && Ecliptic && Galactic && Solar Elong.\\
Field & \colhead{Time} & \colhead{Altitude} & \colhead{$RA$} & \colhead{$Dec$} & 
\colhead{$\lambda$} & \colhead{$\beta$} &
 \colhead{$l$} & \colhead{$b$} & \colhead{$\epsilon$} \\
& \colhead{[s]} & \colhead{[km]} & \colhead{[deg]} & \colhead{[deg]} &
 \colhead{[deg]} & \colhead{[deg]} & \colhead{[deg]} & \colhead{[deg]} & \colhead{[deg]} \\
}
\startdata
(2nd flight \#36.265)&&&&&&&&\\
Vega & 88--98 & 152--168 & 279.24 & 41.28 & 286.10 & 64.20 & 69.95 & 20.06 & 115.77\\
SWIRE/ELAIS-N1 & 109--190 & 185--283 & 242.84 & 54.65 & 209.24 & 72.38 & 84.39 & 44.67 & 93.19\\
NEP & 210--282 & 299--329 & 270.87 & 66.45 & 342.70 & 89.64 & 96.26 & 29.46 & 90.22\\
Elat10 & 305--316 & 328--326 & 226.97 & -2.55 & 225.25 & 14.46 & 356.29 & 45.71 & 115.69\\
Elat30 & 328--348 & 323--315 & 220.86 & 19.80 & 211.14 & 33.75 & 23.00 & 63.36 & 100.35\\
Bootes-A & 358--432 & 310--243 & 218.51 & 34.75 & 200.68 & 46.59 & 58.42 & 66.79 & 91.39\\
Bootes-B & 437--517 & 237--105 & 217.30 & 33.25 & 200.36 & 44.80 & 55.08 & 68.06 & 91.21\\
&&&&&&&&\\
(3rd flight \#36.277)&&&&&&&&\\
Lockman Hole & 125--172 & 202--265 & 161.43 & 58.21 & 135.42 & 45.49 & 149.41 & 51.97 & 118.83\\
SWIRE/ELAIS-N1 & 194--239 & 284--315 & 242.81 & 54.59 & 209.32 & 72.32 & 84.31 & 44.71 & 105.65\\
NEP & 256--309 & 320--324 & 270.63 & 66.28 & 311.48 & 89.62 & 96.06 & 29.56 & 89.76\\
Elat30 & 330--356 & 319--306 & 236.98 & 9.57 & 234.38 & 29.25 & 18.64 & 44.89 & 124.12\\
Bootes-B & 375--425 & 296--244 & 217.30 & 33.27 & 200.35 & 44.82 & 55.13 & 68.06 & 132.31\\
Bootes-A & 451--467 & 211--187 & 218.38 & 34.69 & 200.58 & 46.49 & 58.33 & 66.91 & 130.73\\
&&&&&&&&\\
(4th flight \#40.030)&&&&&&&&\\
DGL & 160--225 & 272--401 & 251.99 & 68.76 & 154.03 & 82.84 & 100.27 & 36.19 & 88.59\\
NEP & 240--300 & 425--505 & 270.80 & 66.15 & 307.86 & 89.48 & 95.91 & 29.49 & 90.32\\
Lockman Hole & 315--365 & 520--558 & 161.23 & 58.58 & 135.01 & 45.74 & 149.07 & 51.65 & 69.33\\
Elat10 & 380--430 & 566--577 & 190.46 & 8.12 & 186.37 & 11.6 & 295.67 & 70.86 & 110.52\\
Elat30 & 445--495 & 577--562 & 193.08 & 28.03 & 179.76 & 30.70 & 110.75 & 89.08 & 102.31\\
Bootes-B & 508--563 & 555--509 & 217.24 & 33.28 & 200.28 & 44.80 & 55.18 & 68.11 & 113.94\\
SWIRE/ELAIS-N1 & 641--696 & 395--275 & 242.76 & 54.68 & 209.05 & 72.38 & 84.45 & 44.71 & 102.07\\
\enddata
\end{deluxetable}

For all the {\it CIBER} experiments, most of the observation time was devoted to observing the core science fields, SWIRE/ELAIS-N1, NEP, and Bootes-A/B, where the {\it Spitzer} and {\it AKARI} satellites carried out deep galaxy surveys and relevant ancillary data from ground-based telescopes are available \citep{2013ApJS..207...32B}.  In order to investigate the ZL foreground, we observed the skies at low ecliptic latitudes, Elat-10 and Elat-30, where the ZL dominates the sky brightness.  The Lockman hole, the lowest cirrus region, was also observed in the third and fourth flights, and a relatively high-cirrus region to investigate the diffuse Galactic light (DGL) was observed in the fourth flight.  However, these additional fields were also useful for airglow investigation, because these observations were done at low altitudes in the early phase of the flight.

To minimize contamination by atmospheric emission and sun light, we set the launch date and time window so that the earth and sun avoidance angles of the science fields were greater than 40 and 115 degrees, respectively.  Moreover, in order to compare our data directly with the {\it COBE/DIRBE} data with little uncertainty, we added another constraint on the solar elongations to be in the range covered by the {\it COBE/DIRBE} observations, 64--128 degrees \citep{1998ApJ...508...25H}.

The actual coordinates of the observation were slightly displaced from the targeted coordinates of the rocket's attitude control system, due to a modest misalignment between it and the LRS telescope. The differences are at most 0.5 degrees or one tenth of the field-of-view (FOV) of the {\it LRS} instrument. The pointing stability is better than 10 arcsec for an exposure time of 50 seconds, which is negligibly small compared to the spatial resolution of the LRS (84 arcsec). Accurate image registration using known 2MASS stars has been performed \citep{2017AJ....153...84K}. The central LRS coordinates of each field are given in Table~\ref{tbl-1}.

\section{Data reduction \& analysis}
\subsection{Image processing}
\begin{figure}
\epsscale{0.7}
\plotone{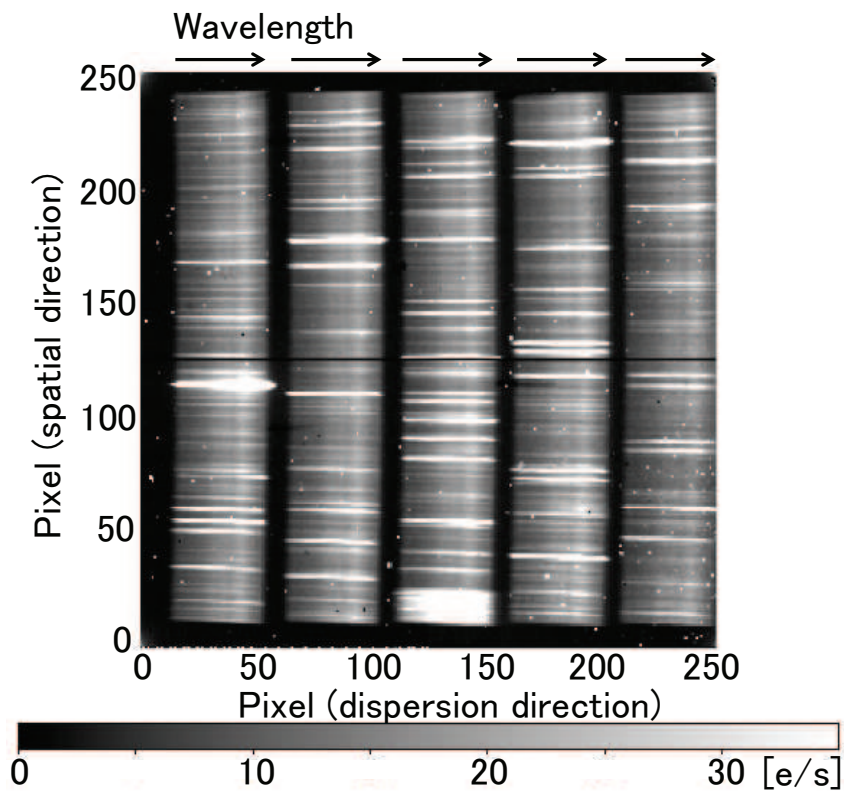}
\caption{An example of a spectral image taken by the LRS at the NEP field in the second flight.  Each of the five columns of the image corresponds to each of the slits.  The wavelength dispersion direction is horizontal and is duplicated across each of the five slits, and the spatial direction (corresponding to the slit length) is vertical.  The grey scale indicates the photocurrent.  The horizontal lines in the image are the dispersed images of bright stars.  The airglow features are seen as dim vertical stripes in all slits.  Bright region at the center bottom corresponds to a square cutout (10$\times$10 pixels) for the PSF measurement in the laboratory.  Randomly distributed bright spots are hot pixels.\label{fig1}}
\end{figure}

The LRS instrument employs a 256$\times$256 HgCdTe focal plane array (FPA) to record the intensity of the dispersed light from a prism through five slits. Figure~\ref{fig1} is an example of the dispersed light image on the FPA with the photocurrent signal normalized to the peak intensity. Sky spectra are calculated from such spectral images using the following process. 

The signal output of the FPA is sampled with multiple reads, recorded for each observation at a frame rate of 4 Hz. The intensity of the image in Figure~\ref{fig1} is directly proportional to the detector signal current, which is calculated from the slope of the accumulated charge in each pixel by line fits to time series data. In this image, stars are easily distinguished as bright spots dispersed horizontally.  We masked all the detector pixels illuminated by stars brighter than about 12 mag in $J-$band and measured the average spectra of the diffuse sky emission using unmasked pixels.  The star mask for each slit was generated by 2$\sigma$ clipping each row after summing the row pixels to improve the signal-to-noise.  The source extraction is complete for stars brighter than 12 mag \citep{2015ApJ...806...69A}.  In this analysis we also exclude the data from the centermost of the five slits as it exhibits anomalous behavior in the early part of the charge integration ramp, which arises from a transient behavior of the readout circuit at corresponding array pixels.

The spectrometer leaves a border of dark pixels that we use to continuously monitor the dark current. To obtain the net photocurrent, the dark current estimated from the masked area is subtracted from the illuminated area of the FOV using an image scaling from shutter-closed data. The dark current difference between the illuminated area and the masked area is estimated to be $\sim$0.03 e$^{-}$s$^{-1}$, which corresponds to a negligible surface brightness of $\sim$0.7 nWm$^{-2}$sr$^{-1}$ \citep{2015ApJ...806...69A}.

\subsection{Generation of sky spectra}
\begin{figure}
\epsscale{0.7}
\plotone{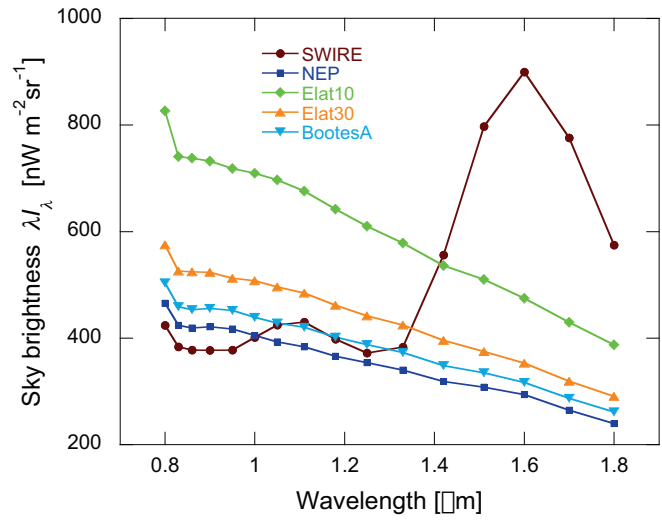}
\caption{Raw sky spectra observed in the second flight.  The sky spectrum at the first field, SWIRE, shows broad peaks at 1.1 and 1.6 $\mu$m due to airglow emission.  The other spectra are similar to each other, without a noticeable 1.6 $\mu$m feature indicating little airglow contamination.\label{fig2}}
\end{figure}

The photocurrent spectrum for each slit is calculated by averaging all the photocurrent signals in spectral columns, the width of each bin corresponding to the spectral resolution, after removing the pixels of bright stars. The dispersion relation between pixel position and wavelength was measured in the laboratory \citep{2013ApJS..207...33T}.  The photocurrent spectra are converted to sky brightness units by a conversion factor for each wavelength determined by a calibration with a diffuse light source in the laboratory \citep{2013ApJS..207...31Z,2015ApJ...806...69A}.  Finally, the sky spectra for the four outer slits are averaged to obtain the final sky spectra.  The final spectral binning from the native wavelength sampling of each pixel is performed so that each spectral bandwidth is equivalent to the spectral resolution, which is determined by the slit-apodized PSF and the prism dispersion, undersampling the spectrum, but reducing correlation between data points.  Sky spectra from the second flight are shown in Figure~\ref{fig2}.  

\subsection{Environmental emission}
\subsubsection{Thermal emission from the rocket skin}
During flights we observe scattered thermal emission from the rocket skin corresponding to a temperature of $T\sim$400 K \citep{2013ApJS..207...31Z}.  This thermal emission was clearly observed in the first flight, and its origin was identified in laboratory tests \citep{2013ApJS..207...33T}.  For the second to fourth flights, the thermal emission was greatly reduced by improving the baffle performance, but still significant at wavelengths longer than 1.8 $\mu$m.  For this work we used only unaffected data at wavelengths shorter than 1.7 $\mu$m, at which the brightness of the thermal emission is estimated from the blackbody fit to the data to be lower than 2 nWm$^{-2}$sr$^{-1}$, similar to the statistical error of the EBL measurement described later.

\subsubsection{Airglow emission}
\begin{figure}
\epsscale{0.7}
\plotone{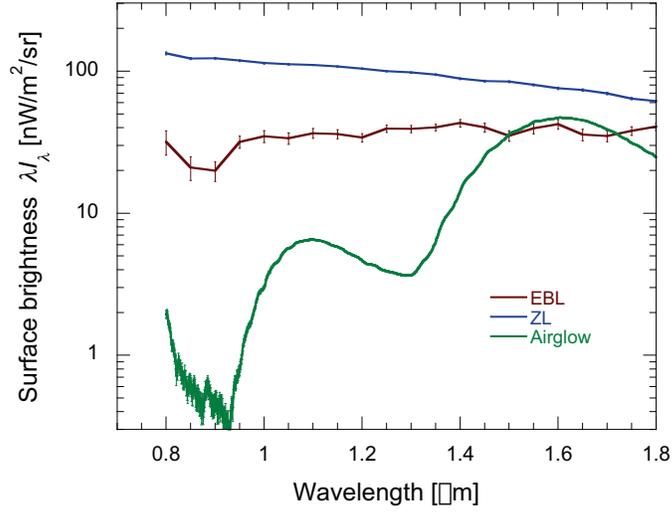}
\caption{Observed airglow spectrum in the second flight.  The airglow emission spectra are derived as a difference of the spectra of science fields taken at different times and altitudes.  The green line is the difference spectrum for the initial field observed early in the flight, showing a clear time-dependent airglow component.  Brightness of the airglow emission components are scaled and compared to the ZL template spectrum (blue line) and the isotropic background spectrum (red line; nominal EBL spectrum presented in the text).\label{fig3}}
\end{figure}

We observe emission from atmospheric airglow and outgassing from the payload at altitudes lower than 250 km.  As detailed in Figure~\ref{fig3}, the emission spectrum of airglow was derived by splitting a field observation into two measurements, and taking the difference to isolate a time- or altitude-dependent signal.  The initial field early in the observation sequence of each flight, \textit{e.g.}, the SWIRE/ELAIS-N1 field for the second flight, shows a strong broad peak around 1.6 $\mu$m, which is plausibly due to vibration transitions of OH molecules smoothed by the spectral resolution of the LRS.  It should be noted that the airglow spectra are quite different from the spectra of either the isotropic background or the ZL to be presented in later sections.  

We modeled the airglow emission based on the time and altitude dependence of the intensity in the fourth flight and a template of the emission spectrum, as described in Appendix A.  For the second flight data, the difference between the spectra including and excluding a slow-decay airglow component gives a conservative estimate for the airglow systematic error. For the fourth flight data, the modeling error of the airglow is accounted in the statistical error of the sky brightness.

\subsection{Measured sky spectra}
\begin{figure}
\epsscale{0.7}
\plotone{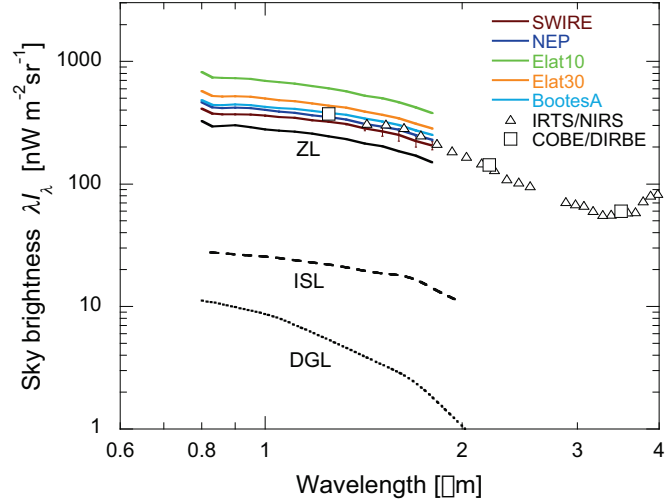}
\caption{Observed near-infrared sky brightness.  Colored lines indicate {\it CIBER} data from the second flight, open squares indicate the darkest {\it COBE/DIRBE} data near the north ecliptic pole \citep{1998ApJ...508...25H}, and open triangles indicate the darkest IRTS/NIRS data at high ecliptic latitudes of 72--73 degrees \citep{2005ApJ...626...31M}.  The combination of statistical error and the systematic error from subtracting atmospheric emission is negligible in the total sky brightness on the scale of this plot.  Absolute calibration errors (not shown) are $\pm3\%$ for {\it CIBER}, $\pm5\%$ for {\it IRTS}, and $\pm1.6\%$ for {\it COBE/DIRBE}.  The thick solid black curve denotes the ZL spectrum derived from the difference among the observed fields, and the thick dashed and thick dotted curves denote ISL and DGL, respectively.  The ISL brightness shown here is calculated for stars of $J >13$ mag.  These foreground amplitudes are indicated for the SWIRE field as typical examples.\label{fig4}}
\end{figure}

Figure~\ref{fig4} shows the measured diffuse sky spectra of the five science fields after masking detected stars.  Foreground emission must be subtracted from the measured absolute sky brightness, including the integrated star light (ISL), DGL, and ZL.  The measured sky levels are similar to the photometric data of {\it COBE/DIRBE} instrument and {\it IRTS} in low-foreground regions\citep{1998ApJ...508...44K,2005ApJ...626...31M}, but with deeper point-source removal due to CIBER's higher angular resolution.  The absolute photometric error of the sky brightness is described in the following sections (see Arai et al. 2015 for further details).

\section{Photometric calibration}\label{cal}
\subsection{Calibration in the laboratory}
We calibrated the LRS response to surface brightness in the laboratory before and after the flights using a range of light sources coupled to an integrating sphere to generate diffuse light.  The radiance from the sphere was absolutely calibrated using an absolutely calibrated spectrograph provided by the National Institute of Standards and Technology \citep{2013ApJS..207...31Z,2013ApJS..207...33T,2015ApJ...806...69A}.  The spectral calibration was performed using a tunable laser and a monochrometer with an accuracy of $\sim$1 nm for all four flights \citep{2015ApJ...806...69A}.  The calibration factor, which is the conversion between the mean pixel photocurrent and surface brightness at each wavelength, is consistent within a 3\% rms variation for the second and third flights, which we adopt as the overall uncertainty of the surface brightness calibration \citep{2015ApJ...806...69A}.

\subsection{In-flight calibration from stars}
The pre-flight calibration of the surface brightness for the fourth flight does not match the calibration determined for the second and third flights\footnote{Though the third flight employed a polarizing grid over four of the five slits, the central slit remained clear and measurements derived from it can be directly compared against the second and fourth flight calibrations.}.  It is reasonable to assume that the calibration does not change much through all the flights, because the optical system is common and there is no evidence that the intrinsic responsivity of PICNIC detectors changes appreciably with time.  It appears that there was some failure in the calibration system used for the pre-flight calibration for the fourth flight, a situation which we did not identify until after the fourth flight, which did not allow payload recovery due to the higher altitude.  To utilize the fourth flight data, we apply the previous calibration and verify that the LRS-observed fluxes of {\it 2MASS} stars are consistent with the catalogued fluxes as described below.

Though the rocket's attitude control system (ACS) aligns the LRS pointing to within $15^{\prime}$ of each target, we require a finer understanding of the instrument pointing to proceed.  We match bright detected stars to the {\it 2MASS} catalogue \citep{2006AJ....131.1163S} to generate the astrometric registration.  To avoid mismatches arising from source confusion, we exclude stars fainter than 10 mag in $J$-band, stars that are next to brighter {\it 2MASS} stars within 500 arcsec, stars that have negative $J$--$H$ color, variable stars, and stars that have non-continuous spectra that could be due to instrumental and/or source confusion noise.  Following these cuts, 105 LRS-detected sources across all fields in all flights are cross-matched with stars appearing in the 2MASS catalog.  Details of the astrometric registration and star spectra are described in a separate paper \citep{2017AJ....153...84K}.

\begin{figure}
\epsscale{0.7}
\plotone{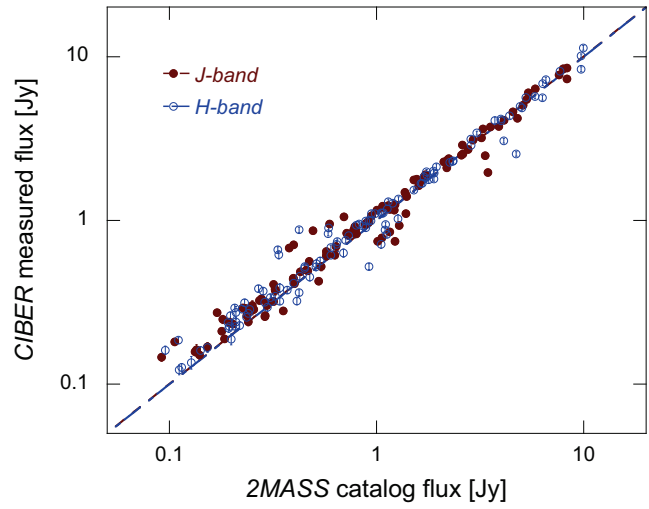}
\caption{Comparison between the {\it 2MASS} cataloged flux and the {\it CIBER} measured flux of bright stars in an equivalent band to the {\it 2MASS} band, for the second, third and fourth flights \citep{2017AJ....153...84K}.  The slit apodization effect is corrected for the {\it CIBER} flux.  Filled circles and open circles indicate the fluxes from all flights in $J$ and $H$ bands, respectively.  The dashed and dash-dotted lines are the results of linear fits to the $J$ and $H$ band data, respectively.\label{fig5}}
\end{figure}

A comparison between the star fluxes from the {\it 2MASS} catalog and the {\it CIBER} observations in all flights is shown in Figure~\ref{fig5}.  A tight correlation, with a slope consistent with unity and with an offset consistent with zero, are apparent in both the $J$ and $H$ bands after correction of the slit apodization.  This indicates that most of the flux detected by the LRS is identical to the flux predicted from the {\it 2MASS} catalog.  Mean ratios of the LRS measured fluxes to the {\it 2MASS} catalog fluxes in $J$ and $H$ bands are 0.45 and 0.46, respectively, and the difference from unity can be explained by the apodizing effect of the slit mask \citep{2017AJ....153...84K}.  Our Monte-Carlo simulation of the stellar photometry, using actual slit function of the LRS and taking the source confusion and instrumental noise into account, predicts the flux range of LRS measured fluxes well \citep{2017AJ....153...84K}.  The measured slope between {\it CIBER} and {\it 2MASS} is 1.01(1.04) for the second and third flights, and 1.03(1.01) for the fourth flight at $J$($H$) band, validating our assumptions for this flight's calibration.

We could also check the calibration over the entire spectral band of the LRS by star observations.  Several stars were accidentally observed in common over more than one flight, and the spectra showed identical spectral shape within the measurement error, indicating consistency of the calibrations among all flights \citep{2017AJ....153...84K}.

\section{Astrophysical foreground analysis}
\subsection{Integrated star light}\label{isl}
\begin{figure}
\epsscale{0.7}
\plotone{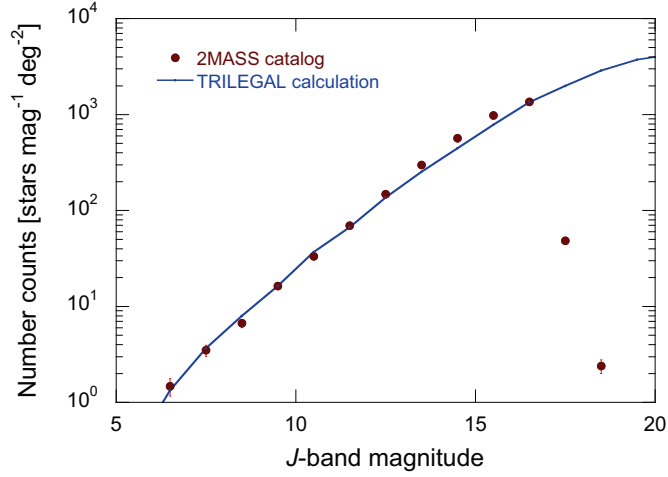}
\caption{Comparison of the {\it 2MASS} counts and the model star counts in $J$-band at NEP.  The {\it 2MASS} counts (filled circles) are well reproduced by the model star counts (thin line) at magnitudes below the detection limit of the LRS of $J <$13 mag.  Results in the other fields are similar.\label{fig6}}
\end{figure}

\begin{figure}
\epsscale{0.5}
\plotone{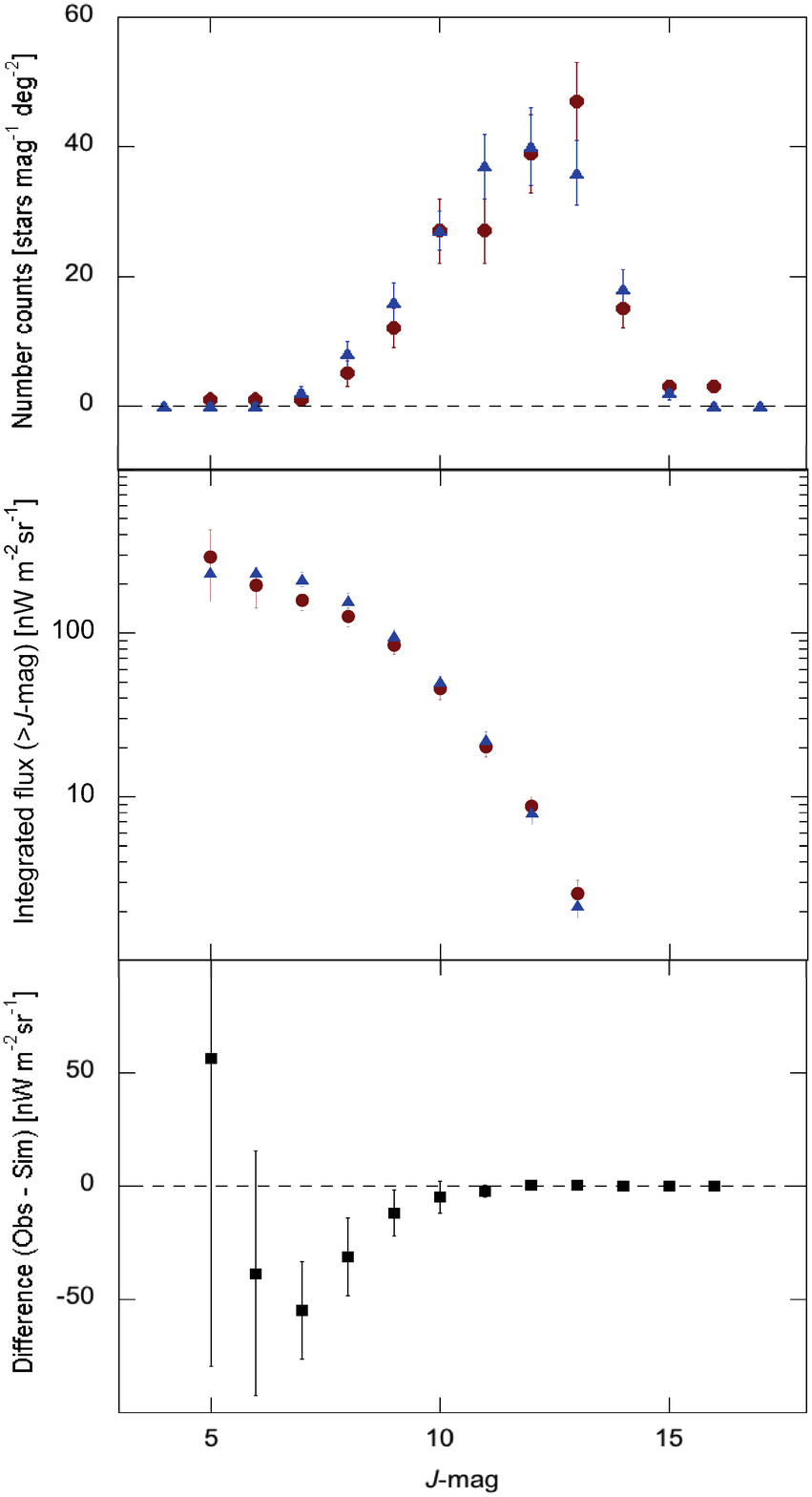}
\caption{Comparison between the LRS-observed star counts and simulated star counts.  Top: The LRS counts in $J$-band at NEP (circles) agree with the simulation (triangles) to within Poisson errors.  Middle: Same comparison on the surface brightness of the integrated flux of stars fainter than the given magnitude.  Bottom: The difference between the observation and the simulation is close to zero at the magnitude range with the best statistics, i.e., $J>$9, which is relevant to the residual ISL for the LRS after the $J<$12 cut.\label{fig7}}
\end{figure}

To estimate the surface brightness of the ISL, we begin with the $J$, $H$ and $K$ band flux of the {\it 2MASS} sources \citep{2006AJ....131.1163S} within the LRS's field of view (FOV), taking into account the survey completeness and the effective slit width \citep{2015ApJ...806...69A}.  The FOV of the LRS is determined from the slit width and the point spread function measured in the laboratory using monochromatic light sources \citep{2013ApJS..207...31Z,2013ApJS..207...33T,2015ApJ...806...69A}.  To simulate the completeness magnitude and calculate the continuous spectrum of ISL, we carried out additional simulations by the same method, but using a model of star counts based on stellar population synthesis, TRILEGAL \citep{2005A&A...436..895G}.  The simulated counts of bright stars in the {\it 2MASS} bands agree with the {\it 2MASS} counts up to the detection limit of the LRS ($J\sim$13 mag) as shown for the NEP field in Figure~\ref{fig6}, and with our observed star counts within Poisson errors as shown in Figure~\ref{fig7}.  The results in the other fields are similar.  We also confirm that the ISL spectrum from the star count model is consistent with the $JHK$ color of the ISL from {\it 2MASS} for 9 $< J <$ 17 to within 2\% accuracy.

\begin{figure}
\epsscale{0.7}
\plotone{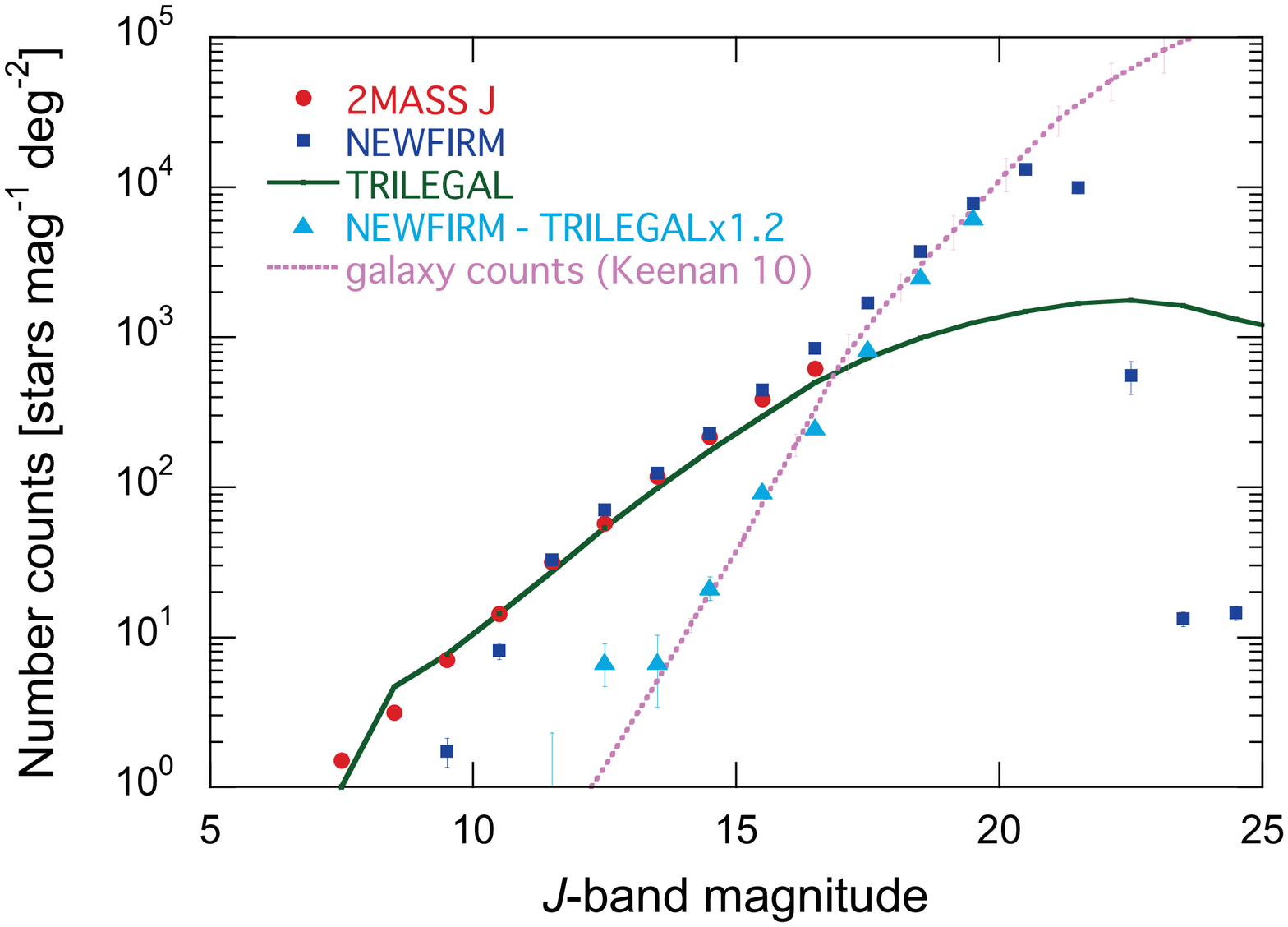}
\caption{Comparison of the model star counts with deep observations toward the Bootes field in $J$-band.  The {\it 2MASS} counts (red circles) are reproduced by the TRILEGAL star count model (green solid line) \citep{2005A&A...436..895G} below the limiting magnitude of the LRS $J\sim$13 mag.  A slight enhancement (about 20\%) of the TRILEGAL model counts is required to fit the {\it 2MASS} counts at fainter magnitudes.  The galaxy counts (light blue triangles) estimated from the difference between the observed total counts from the NEWFIRM observation in the Bootes field (blue squares)  \citep{2010AAS...21641513G} and the TRILEGAL model scaled by about 20\% are consistent with well-established galaxy counts (pink dotted line) \citep{2010ApJ...723...40K}.\label{fig8}}
\end{figure}

We extend the ISL calculation to magnitudes fainter than the limiting magnitude of {\it 2MASS} $AB$=17 mag using the same star count model, but scaled by a factor of 1.2 to fit the {\it 2MASS} counts in the magnitude range from $J$=13 to 17 mag.  The validity of the $\times$1.2 scaled model at fainter magnitudes than 17 mag was checked using deep star counts complete to $AB$=20 mag in $J$-band towards Bootes and NEP with {\it NEWFIRM} \citep{2010AAS...21641513G} and {\it WIRCam/CFHT} \citep{2014A&A...566A..60O}, respectively.  The raw source counts (including galaxies) towards Bootes are compared with the model star counts and also with well-established galaxy count data \citep{2010ApJ...723...40K} as shown in Figure~\ref{fig8}.  We confirm that the sum of the model star counts and the galaxy counts can account for the observed total counts.  In Table~\ref{tbl-2}, the model prediction for the stars fainter than the {\it 2MASS} detection limit is compared with the deep counts for Bootes and NEP.  From this result, the uncertainty of ISL at faint magnitudes is estimated to be 1.5 nWm$^{-2}$sr$^{-1}$ in $J$-band, dominated by uncertainties of the galaxy counts \citep{2011MNRAS.410.2556D} in separating stars from galaxies.

\begin{deluxetable}{llllll}
\tabletypesize{\scriptsize}
\tablecaption{Verification of the ISL estimate using the source counts data\label{tbl-2}}
\tablewidth{0pt}
\tablehead{
\colhead{Field}&\colhead{Total\tablenotemark{a,b}}&
\colhead{IGL\tablenotemark{a}}&\colhead{ISL\tablenotemark{a}}&
\colhead{ISL\tablenotemark{a}}&\colhead{Reference}\\
&\colhead{(1)}&\colhead{(2)}&\colhead{(1)--(2)}&\colhead{Model}&\\
}
\startdata
Bootes & 5.45 & 4.31$\pm$1.02 & 1.14$\pm$1.02  &1.93 & Gonzalez et al. 2010\\
NEP & 9.64 & 4.31$\pm$1.02 & 5.33$\pm$1.02 &5.52 & Oi et al. 2014\\
\enddata
\tablenotetext{a}{Integrated flux for $J >$17 mag in nWm$^{-2}$sr$^{-1}$ at 1.25 $\mu$m.}
\tablenotetext{b}{Total integrated source counts including IGL and ISL.}
\end{deluxetable}

The resulting ISL spectrum at the SWIRE/ELAIS-N1 field is shown in Figure~\ref{fig4}.  The ISL spectra in the other fields are identical to this even though the brightness varies by factor of 3.  The total uncertainties of ISL at our observation fields are estimated to be 5--10\%, which consists of the Poisson error, slight color difference between the star count model and the {\it 2MASS} counts, and uncertainty in the contribution of galaxies at fainter magnitudes \citep{2010ApJ...723...40K}.  The total systematic uncertainty of the ISL estimate propagated to the final EBL estimate is described in Section~\ref{systematics}.

\subsection{Diffuse Galactic light}
The DGL, starlight scattered by interstellar dust, correlates with the far-infrared brightness of interstellar dust emission in optically thin regions \citep{2012ApJ...744..129B}.  The DGL spectrum was previously measured with {\it CIBER} by correlating the brightness over each field with the 100 $\mu$m intensity \citep{2015ApJ...806...69A}.  The field-averaged DGL spectrum exhibits a continuum approximated by a model spectrum as shown in Figure~\ref{fig4}.  The LRS-measured DGL spectrum is consistent with the result in the visible with {\it Pioneer10/11} \citep{2011ApJ...736..119M}, with {\it COBE/DIRBE} \citep{2015ApJ...811...77S}, and with {\it AKARI} at wavelengths longer than 1.8 $\mu$m \citep{2013PASJ...65..120T}.  All the DGL measurements fit a DGL model spectrum from \citet{2004ApJS..152..211Z} with a scaling by a factor of 2 \citep{2015ApJ...806...69A,2015ApJ...811...77S}.  In this work, we use the best-fit scaled Zubko et al. DGL model to estimate the DGL brightness according to the 100 $\mu$m brightness at each science field.  We evaluate the uncertainty of our DGL estimate as 20\% from the model scaling error to the measured DGL brightness, which does not appreciably affect the accuracy of the EBL measurement \citep{2015ApJ...806...69A}.  The propagation of the DGL uncertainty to the EBL estimate is described in Section~\ref{systematics}.

\subsection{Zodiacal light}
\begin{figure}
\epsscale{0.7}
\plotone{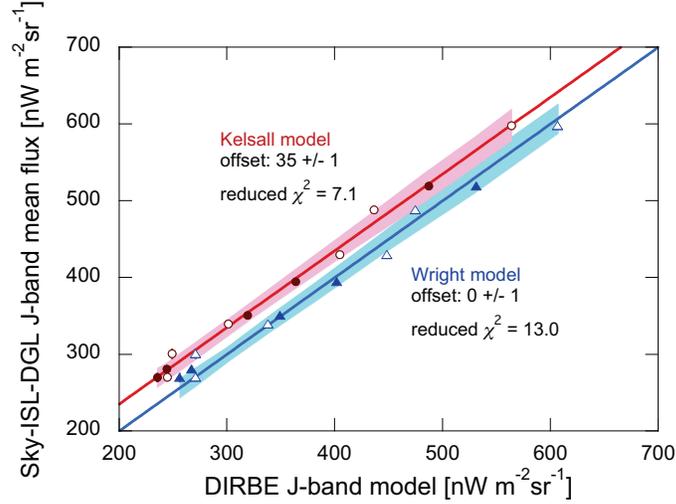}
\caption{The Sky-ISL-DGL brightness compared with the ZL models.  The mean flux of the LRS sky spectra in {\it J}-band (1.25 $\mu$m) after subtracting ISL and DGL foregrounds are plotted against published ZL models.  The filled circles and triangles denote the data from the second flight, and the open circles and triangles denote the data from the fourth flight.  The thick lines are unity slope to fit the data by setting the offsets as free parameters.  The error bar of each data point indicates statistical error.  Systematic uncertainties including the calibration error of 3\%, and the modeling error of ISL and DGL subtraction, which is estimated from the ISL and DGL intensities in each field, are shown by the shaded area.  The Kelsall ZL model \citep{1998ApJ...508...44K} (filled and open circles; $\chi^{2}$/d.o.f.=7.1) shows a better fit to the data than the Wright ZL model  \citep{2001IAUS..204..157W} (filled and open triangles; $\chi^{2}$/d.o.f.=13.0), with $\chi^{2}$ quoted for errors including ISL and DGL subtraction errors.  The Kelsall ZL model is favored by the data, but the Wright ZL model cannot be fully excluded.\label{fig9}}
\end{figure}

To obtain the final EBL spectrum, we subtract a smooth ZL component using two models based on {\it COBE/DIRBE} all-sky measurements \citep{1998ApJ...508...44K,2001IAUS..204..157W}.  To confirm the applicability of the ZL model, we correlate the ISL- and DGL-subtracted surface brightness at {\it DIRBE} $J$-band with the Kelsall ZL model \citep{1998ApJ...508...44K},  shown in Figure~\ref{fig9}.  We observe a tight correlation between the data and the ZL model, consistent with the estimated uncertainties from statistical, systematic, and modeling errors.  We also carry out the same test using the alternative Wright ZL model \citep{2001IAUS..204..157W}.  The Wright ZL model gives a poorer fit to the data, but due to the measurement's systematic errors, this model cannot be excluded.

\begin{figure}
\epsscale{0.8}
\plotone{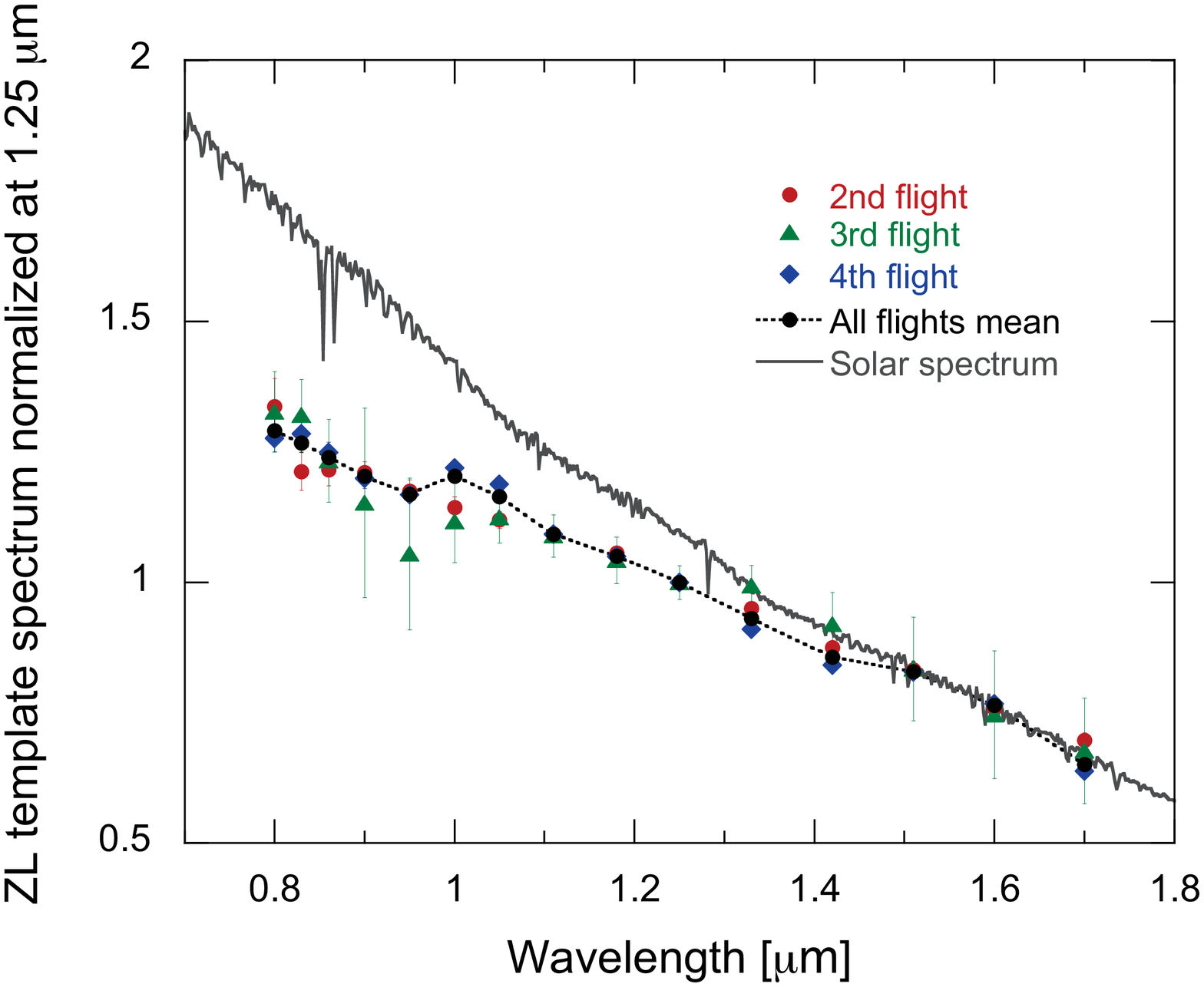}
\caption{\small Template spectrum of ZL.  The weighted mean of difference sky spectra observed in the second, third and fourth flights is indicated by the circles, triangles and diamonds, respectively.  For differing the sky spectra, the Vega and Bootes-B fields are not used from the second flight, the Lockman Hole field is not used from the third flight, and the DGL field is not used from the fourth flight.  The third flight ZL template is shown here for a reference, while the third flight data are not used for the EBL analysis as described in Section~\ref{flights}.  The error bars indicate the weighted standard deviation for the difference spectra.  The filled circles connected with dotted line denote the mean spectrum for the all flights.  A few per cent differences in the ZL templates among the flights exceeding the error bars could be due to systematic error in the gain calibration.  Our measured ZL spectrum is redder than solar spectrum (thin line, \url{http://rredc.nrel.gov/solar/spectra/am0}), and the weak spectral features from 0.9 to 1.4 $\mu$m causing the reddening are associated with silicates as reported by first flight {\it CIBER} measurements \citep{2010ApJ...719..394T}.  The amplitude of the solar spectrum is scaled to the data at 1.6 $\mu$m.\label{fig10}}
\end{figure}

To apply the ZL model calculated at $J$-band to the other LRS wavelengths, we scale a spectral template of the ZL.  This template is computed from the ISL- and DGL-subtracted spectra by forming the pair-wise difference of the spectra, normalizing the spectra, and then computing the mean of all such spectral differences in a given flight with weights calculated from the statistical errors.  The field differencing scheme effectively cancels out isotropic components of the emission.  Most of the data points in the resulting ZL template spectra are consistent with each other within the uncertainties, as shown in Figure~\ref{fig10}.  Some differences between the flights beyond the expected errors may be due to systematic errors in the gain calibration of the LRS.  In this work, therefore, we apply the ZL template measured from a given flight to its associated flight data to avoid increasing the effect of calibration error in the ZL subtraction.  No ZL model uncertainty is explicitly propagated to the derived EBL brightness, though the absolute instrument calibration error is taken into account.  Rather, the ZL modeling error is captured through the comparison of the Kelsall and Wright models.

\section{Extragalactic background result}
\subsection{Nominal EBL spectrum}
\begin{figure}
\epsscale{0.8}
\plotone{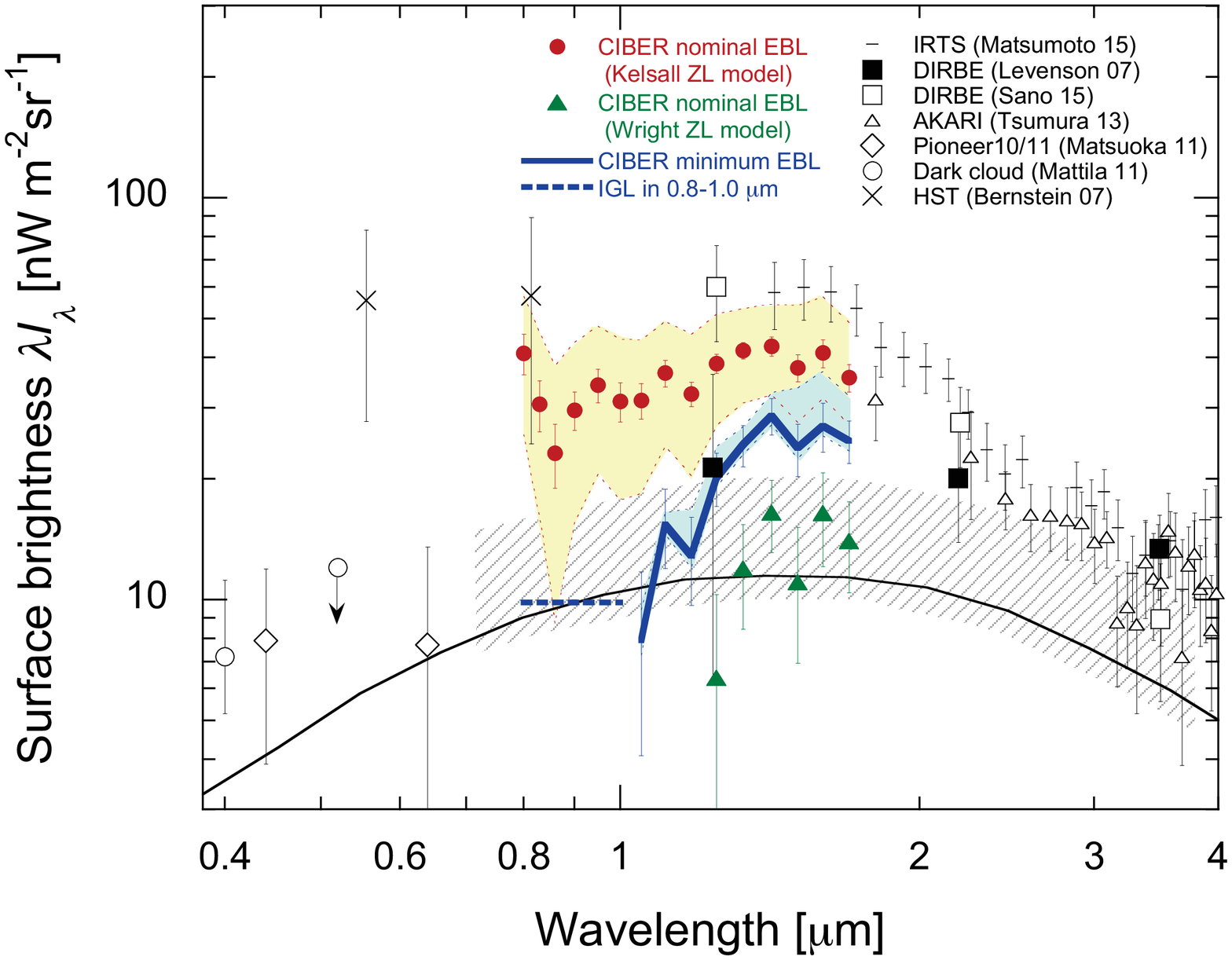}
\caption{\small The measured EBL spectrum.  Our nominal EBL result with the combined data from the two flights assuming the Kelsall ZL model \citep[filled circles]{1998ApJ...508...44K} is compared with previous results by {\it COBE} \citep[filled \& open squares]{2007ApJ...666...34L,2015ApJ...811...77S,2016ApJ...821L..11S}, {\it IRTS} \citep[horizontal bars]{2015ApJ...807...57M} and {\it AKARI} \citep[open triangles]{2013PASJ...65..121T} in the near-infrared, and by {\it Pioneer10/11} \citep[open diamonds]{2011ApJ...736..119M}, {\it HST} \citep[crosses]{2007ApJ...666..663B} and observations from the ground with the dark-cloud method \citep[open circles]{mattila11} in the visible.  Our error bars indicate the total statistical error.  The dotted lines are upper and lower bounds (68\% confidence) on our nominal result for the total systematic error including the absolute calibration error, and the modeling error on ISL and DGL subtraction.  The thin solid curve gives the IGL derived from deep galaxy counts \citep{2010ApJ...723...40K}.  The filled triangles (with statistical errors only) give our measured EBL using the Wright ZL model  \citep{2001IAUS..204..157W}, which produces an unphysical EBL below the IGL at $\lambda <$ 1.3 $\mu$m.  The thick solid line indicates a model-independent lower limit (Minimum EBL) with statistical error bars, and the upper and lower dotted lines are systematic error boundary (see text).  This limit is derived by subtracting a scaled amplitude of ZL such that the derived EBL matches the IGL at 0.8--1.0 $\mu$m, given by the thick dashed line.  The hatched region indicates the error boundary of the EBL derived from intergalactic absorption of gamma rays by HESS \citep{2013A&A...550A...4H}.\label{fig11}}
\end{figure}

The nominal EBL spectrum computed from the combined second and fourth flight data after subtracting the Kelsall ZL model template spectrum is shown in Figure~\ref{fig11} and summarized in Table~\ref{tbl-3}.  The 1$\sigma$ statistical uncertainty is indicated by the error bars on the data points, while the total systematic uncertainty of the nominal EBL spectrum is indicated by an error band.  The nominal near-IR EBL surface brightness is similar to or slightly lower than the {\it DIRBE} \citep[references of the other previous works are therein.]{2001ApJ...555..563C,2015ApJ...811...77S} and {\it IRTS} \citep{2005ApJ...626...31M,2015ApJ...807...57M} measurements derived using the Kelsall ZL model, but higher than the DIRBE measurement derived using the Wright ZL model \citep{1998ApJ...496....1W,2007ApJ...666...34L}. We find a nominal EBL surface brightness of 42.7$^{+11.9}_{-10.6}$ nWm$^{-2}$sr$^{-1}$ at 1.4 $\mu$m.  This is out of allowed range of the EBL of 15$\pm$5 nWm$^{-2}$sr$^{-1}$ at 1.4 $\mu$m derived from {\it HESS} (High Energy Spectroscopic System) gamma-ray absorption spectra \citep{2013A&A...550A...4H}.  Our derived EBL spectrum, which decreases toward visible wavelengths from a peak around 1.5 $\mu$m, cannot be attributed solely to residual ZL foreground, as the spectrum of the EBL is redder than the ZL component shown in Figure~\ref{fig10}.

\begin{deluxetable}{llll}
\tabletypesize{\scriptsize}
\tablecaption{Summary of the EBL results\label{tbl-3}}
\tablewidth{0pt}
\tablehead{
\colhead{Wavelength}&\colhead{Nominal EBL\tablenotemark{a}}&&
\colhead{Minimum EBL\tablenotemark{a}}\\
\colhead{[$\mu$m]}&\colhead{[nWm$^{-2}$sr$^{-1}$]}&&
\colhead{[nWm$^{-2}$sr$^{-1}$]}\\
}
\startdata
0.80 & 41.1$\pm$4.8 ${+15.7}/{-15.3}$ && - \\
0.83 & 30.6$\pm$4.4 ${+15.8}/{-14.7}$ && - \\
0.86 & 23.1$\pm$4.2 ${+15.1}/{-14.4}$ && - \\
0.90 & 29.6$\pm$3.2 ${+14.1}/{-14.0}$ && - \\
0.95 & 34.2$\pm$3.3 ${+13.8}/{-13.7}$ && - \\
1.00 & 31.2$\pm$3.5 ${+13.4}/{-13.5}$ && - \\
1.05 & 31.3$\pm$3.2 ${+13.1}/{-12.9}$ && 7.9$\pm$3.8 ${+1.5}/{-0.6}$\\
1.11 & 36.7$\pm$2.8 ${+12.6}/{-12.6}$ && 15.4$\pm$3.4 ${+1.1}/{-0.8}$\\
1.18 & 32.5$\pm$2.2 ${+13.3}/{-12.2}$ && 12.9$\pm$3.2 ${+4.0}/{-0.9}$\\
1.25 & 38.7$\pm$2.1 ${+12.6}/{-11.6}$ && 20.1$\pm$3.1 ${+3.9}/{-1.1}$\\
1.33 & 41.7$\pm$2.0 ${+11.3}/{-10.9}$ && 24.3$\pm$2.8 ${+2.7}/{-1.3}$\\
1.42 & 42.7$\pm$2.3 ${+11.7}/{-10.3}$ && 28.7$\pm$3.0 ${+4.1}/{-1.5}$\\
1.51 & 37.8$\pm$2.9 ${+16.4}/{-10.1}$ && 23.8$\pm$3.5 ${+9.9}/{-1.4}$\\
1.60 & 41.1$\pm$3.3 ${+15.6}/{-9.3}$ && 27.1$\pm$3.8 ${+10.0}/{-1.5}$\\
1.70 & 35.7$\pm$2.8 ${+13.3}/{-8.4}$ && 24.8$\pm$3.0 ${+7.2}/{-1.4}$\\
\enddata
\tablenotetext{a}{Mean value $\pm$ Statistical error + Systematic error(upper/lower)}
\end{deluxetable}

\subsection{EBL with Wright ZL model}
We must evaluate our nominal result given the large uncertainties of the ZL models.  We also calculate the EBL assuming the Wright ZL model with a "very strong no zodi principle" \citep{2001IAUS..204..157W}, and confirm the EBL is 7.0$^{+13.1}_{-12.6}$ nWm$^{-2}$sr$^{-1}$ in an equivalent DIRBE band at 1.25 $\mu$m, which is consistent with Levenson's et al. EBL of 21.3$\pm$15.1 nWm$^{-2}$sr$^{-1}$ within the errors.  However, the derived EBL is negative for $\lambda$ $<$ 1.0 $\mu$m and less than the IGL for $\lambda$ $<$ 1.3 $\mu$m.  The EBL derived from the Wright model is unphysical, it is also redder than the ZL spectrum.  As previously noted, the Wright model does not fit the observed field difference data as well as the Kelsall model.

\subsection{Minimum EBL spectrum}
It is instructive to consider the EBL spectrum that results from making the most aggressive assumptions about the surface brightnesses of the ZL component.  To obtain a conservative minimum estimate of the 1.0--1.8 $\mu$m EBL, we set the 0.8--1.0 $\mu$m signal levels to the IGL brightness as follows.  We subtracted the ZL spectrum with an arbitrary scaling to eliminate any excess EBL at 0.8--1.0 $\mu$m and to match the IGL from galaxy counts \citep{2011MNRAS.410.2556D}.  The scaling would be interpreted as an isotropic ZL component, but its physical origin is unclear.  

The mean "minimum EBL" spectrum is shown in Figure~\ref{fig11} (also in Table~\ref{tbl-3}), with errors computed from the field-to-field variance and the statistical error of the derived EBL.  The ZL brightness subtracted to obtain the minimum EBL is $\sim$5\% higher than the Kelsall ZL model brightness at high ecliptic latitudes.  Due to the blue color of the ZL spectrum, the derived EBL remains large at longer wavelengths: at 1.4 $\mu$m, the minimum EBL brightness is 28.7$^{+5.1}_{-3.3}$ nWm$^{-2}$sr$^{-1}$.  This exceeds the EBL upper limits from high-energy gamma-ray observations \citep{2013A&A...550A...4H}.  Our result indicates that underestimating the ZL foreground \citep{2005ApJ...635..784D} cannot eliminate an excess above IGL near 1.5 $\mu$m.  If the minimum EBL calculation is the true EBL level, the EBL must have a very red color increasing from 1.0 $\mu$m to 1.7 $\mu$m that is quite different from the IGL spectrum.

\subsection{Field-to-field Variation of the EBL spectra}
\begin{figure}
\epsscale{0.7}
\plotone{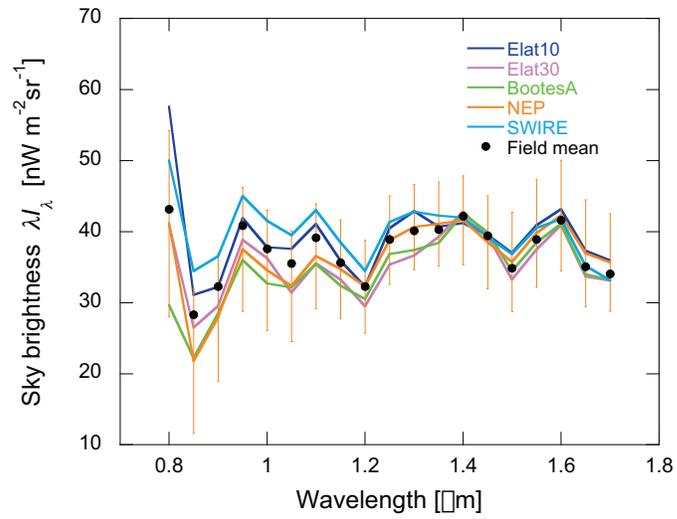}
\caption{Residual EBL spectrum obtained from the second flight by direct subtraction of the Kelsall ZL model at each field.  The total error excluding the absolute calibration error is shown only for the NEP field as a typical example.  The weighted mean of these spectra is shown as filled circles.\label{fig12}}
\end{figure}

The variation of the EBL spectra over the five fields observed in the second flight using the Kelsall ZL model are shown in Figure~\ref{fig12}.  The brightness difference between the fields is only a few nWm$^{-2}$sr$^{-1}$ at most wavelengths, except at short wavelengths which have relatively large statistical errors.  The uncertainty-weighted mean of the spectra is adopted as the nominal EBL result, and the field-to-field variation is accounted as the statistical error of the final EBL result shown in Figure~\ref{fig11}.

\subsection{Systematic uncertainties of the EBL result}\label{systematics}
\subsubsection{Airglow subtraction uncertainties}
As described in Appendix~\ref{airglow}, we estimate airglow contamination using a double-exponential function.  The uncertainty in the fit parameters is propagated through the analysis and is accounted as part of the total statistical uncertainty of the sky brightness measurement.  Although we subtract both fast and slow decay airglow components from the second flight data using a model based on the fourth flight data, the evidence that the second flight data require a slowly decaying component is weak.  We account for this possibility as the major systematic error in the airglow subtraction; difference between the EBL estimates with and without subtraction of the slow airglow component.

\subsubsection{ISL and DGL Uncertainties}
\begin{figure}
\epsscale{0.7}
\plotone{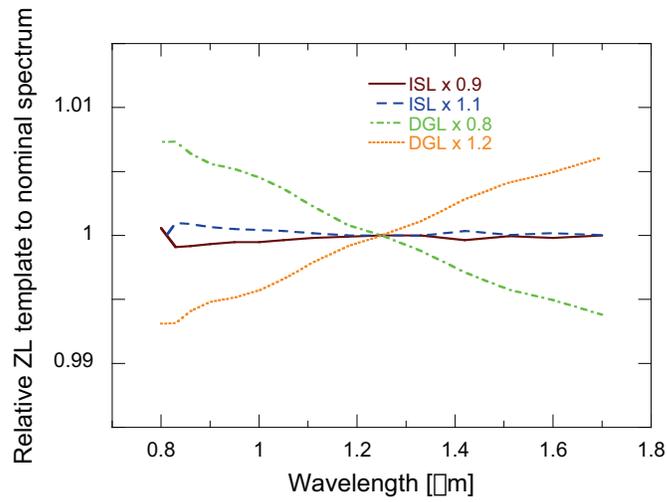}
\caption{Dependence of ZL template spectra on ISL and DGL intensities.  ZL template spectra are calculated for the ISL intensity increased (solid line) and decreased (dashed line) by 10\% from the nominal case and for the DGL intensity increased (dot-dashed line) and decreased (dotted line) by 20\% from the nominal case.  Relative colors after dividing those spectra by the nominal ZL template spectrum are shown.\label{fig13}}
\end{figure}

\begin{figure}
\epsscale{1.0}
\plottwo{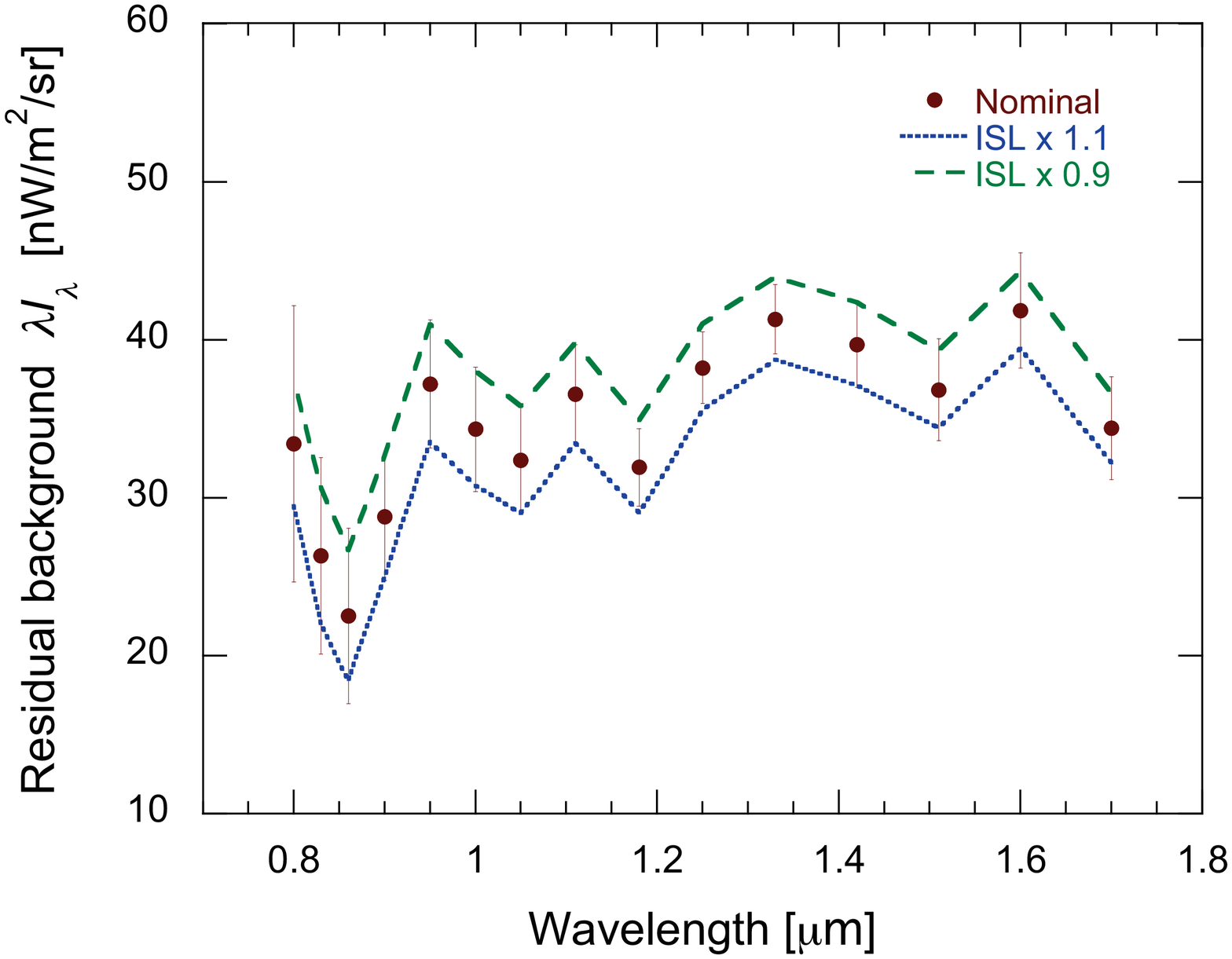}{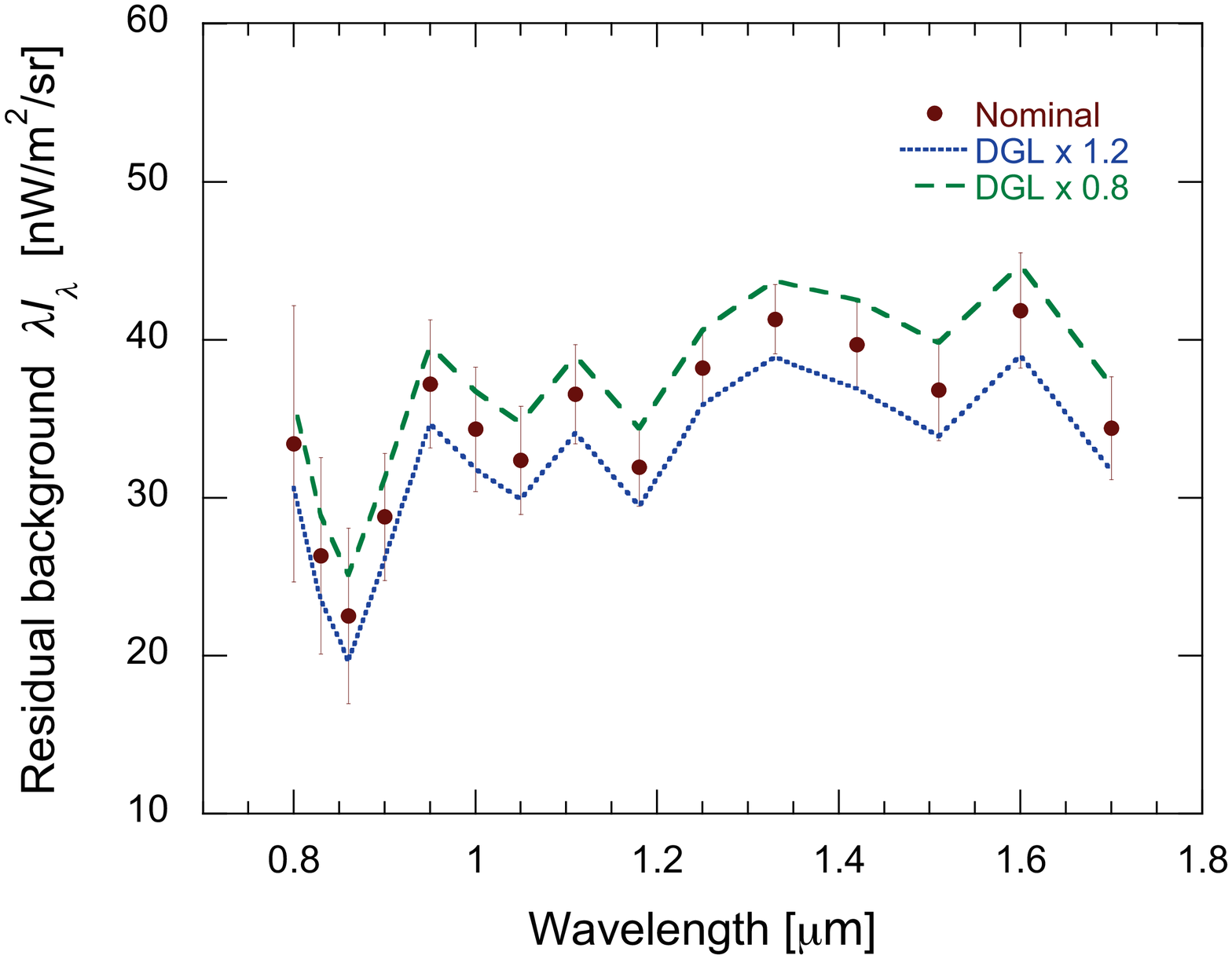}
\caption{{\it Left}: Susceptibility of the EBL estimate to a change in ISL intensity.  The residual background calculated for the ISL increased (dashed line) and decreased (dotted line) by the absolute ISL uncertainty of 10\% are compared with the nominal EBL result from the second flight (filled circles) with statistical error.  {\it Right}: Same as the left panel but for the DGL increased and decreased by the absolute DGL uncertainty of 20\%.\label{fig14}}
\end{figure}

We determine the susceptibility of the EBL measurement to differences in the assumed ISL by changing the ISL amplitude by $\pm$10\% from its nominal value.  The ZL template spectrum is affected as shown in Figure~\ref{fig13}.  The total effect on the EBL estimate, including the change from the modified ZL template, is shown in the left panel of Figure~\ref{fig14}.  We estimate the ISL intensity error from uncertainties in the galaxy counts described in Section~\ref{isl} to be 10\%.  The spectral uncertainty of the ISL is negligible compared to the absolute intensity error.  We assign a systematic error of 5 nWm$^{-2}$sr$^{-1}$ at 1.25 $\mu$m from the uncertainties associated with ISL.

In a similar fashion, we estimate the error in the EBL measurement from the DGL model by varying the DGL brightness $\pm$20\% from its nominal value, as shown in the right panel of Figure~\ref{fig14}.  Again, the effect of the DGL on the ZL template spectrum is propagated through to the EBL estimate, as shown in Figure~\ref{fig13}.  In this case, the systematic error from DGL propagated to the mean EBL brightness is approximately 4 nWm$^{-2}$sr$^{-1}$ at 1.25 $\mu$m.

\subsubsection{Comparison of the EBL Results between Different Flights}
EBL measurements from different flight data should give the same results, though data quality and systematic errors do vary from flight to flight.  As discussed in Section~\ref{cal}, since the fourth flight laboratory calibration is not reliable, we use the third flight calibration factor for the fourth flight data.  Note that the difference between the calibration factors of the second and third flights is only 1\% in the entire LRS bands.

\begin{figure}
\epsscale{1.0}
\plottwo{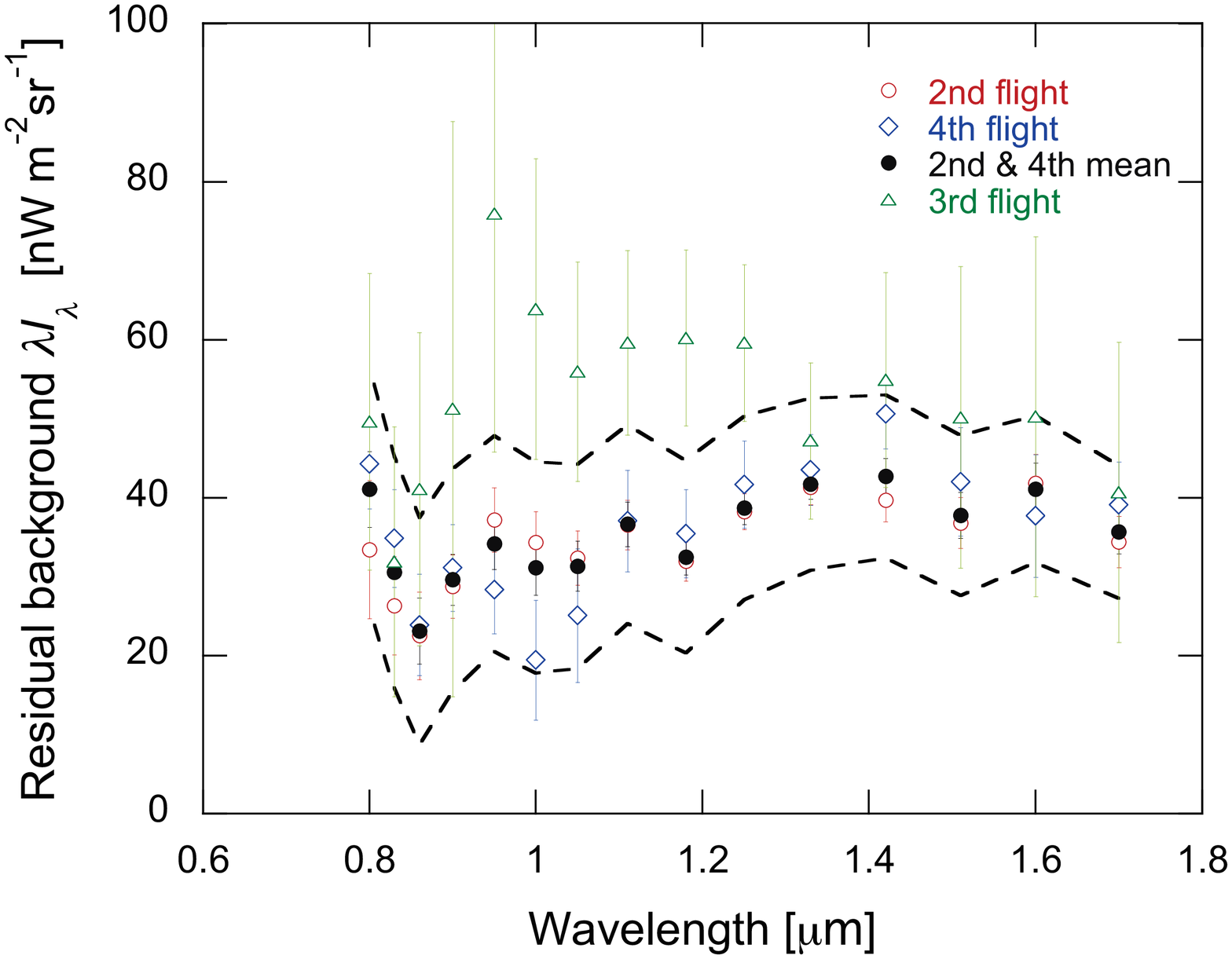}{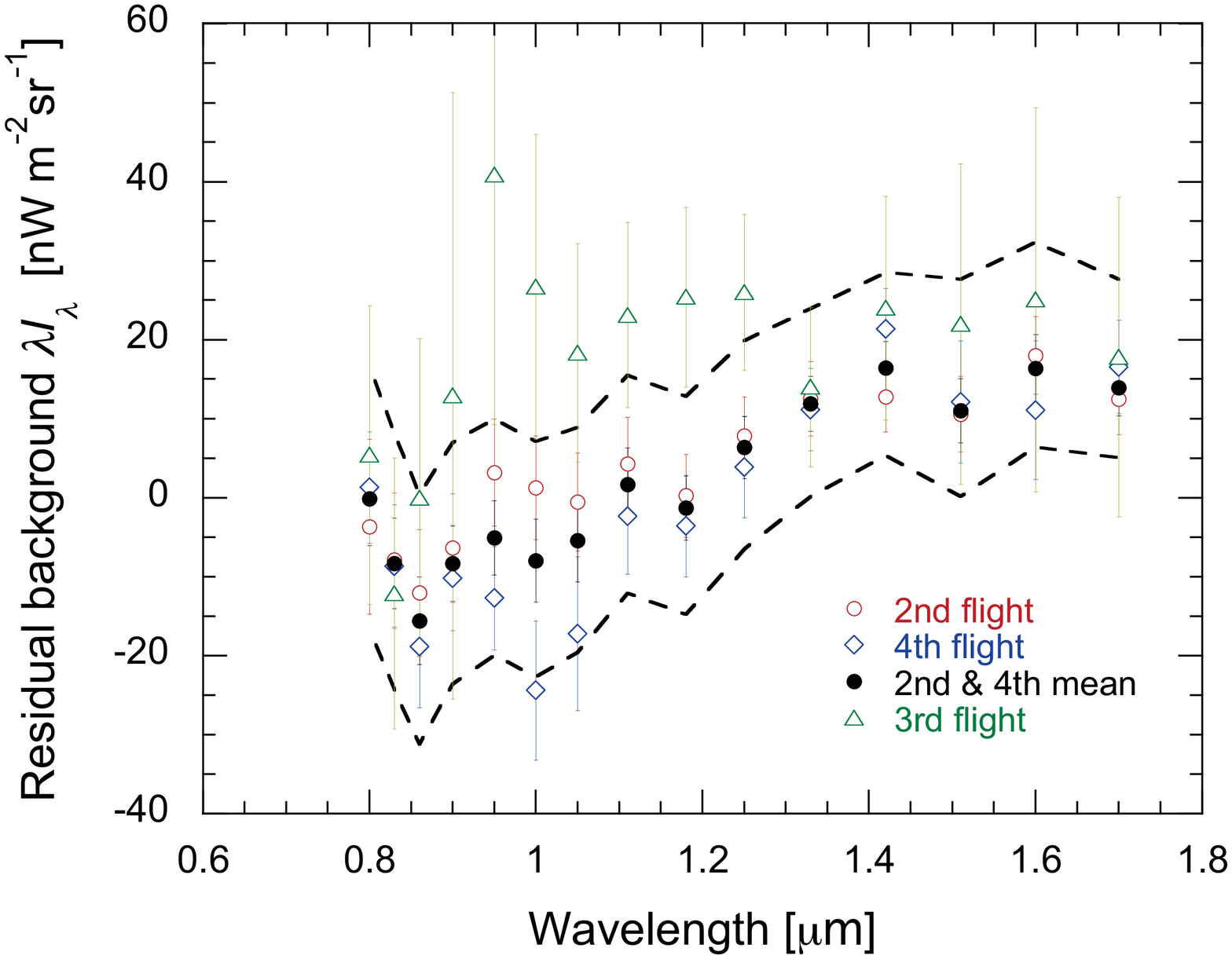}
\caption{{\it Left}: The measured EBL with the Kelsall ZL model for all flights.  Mean residual background spectra obtained by direct subtraction of the Kelsall ZL model are shown; open circles - second flight, open diamonds - fourth flight, open triangles - third flight, and filled circles - second and fourth combined result shown in Figure~\ref{fig11}.  The error bars are total statistical errors including the field-to-field variations.  The dashed lines give upper and lower boundaries on the systematic errors of the second and fourth combined result.  {\it Right}: Same as the left panel but with the Wright ZL model.\label{fig15}}
\end{figure}

We compare the EBL results for each of the three flights with one another In Figure~\ref{fig15}.  As we expected, the fourth flight result agrees with the second flight result within the error bars, except for a dip in the fourth flight data at 1 $\mu$m, where the measurement error is large due to presence of the airglow systematic (Appendix~\ref{slowag}).  The signal-to-noise ratio of the third flight data is comparably smaller because only the non-polarized channel, which corresponds to one of five slits on focal plane mask, is available for measuring the total brightness.  The resulting spectrum has errors a factor of $2.2$ larger than the others, and we observe a large discrepancy from the other flights data near 1 $\mu$m.

\subsubsection{Total Error Budget}
The total error budget of our EBL measurement referenced to 1.25 $\mu$m is enumerated in Table~\ref{tbl-4}.  The statistical and systematic errors of the EBL estimate at other wavelengths are shown graphically in Figure~\ref{fig11}.  The ZL model error is not included in the final EBL error, but is bracketed by the difference between the two published ZL models, as shown in Figure~\ref{fig15}.  The different sources of statistical and systematic errors that arose in the data reduction and foreground subtraction are summed quadratically to produce the final EBL errors.

\begin{deluxetable}{lll}
\tabletypesize{\scriptsize}
%\rotate
\tablecaption{Total Error Budget for the EBL measurement\label{tbl-4}}
\tablewidth{0pt}
\tablehead{
Error term & Error\tablenotemark{a} & Comment
}
\startdata
Raw sky & $\pm$1 & Detector noise, dark current subtraction\\
& 3\% ($\pm$12 at NEP) & Absolute calibration error\\
Airglow & $<$1 & Model fitting error, maximum at SWIRE\\
& $+$4 & In case of no subtraction of the slow airglow component\\
ZL template & 1\% & Variation of sky difference spectra\\
& 0.4\% & ISL and DGL modeling error\\
& (1\%)\tablenotemark{b} & Spectral response calibration error\\
ISL & 10\% ($\pm$5 at NEP) & Estimated from simulation (Poisson noise, aperture correction,\\
&& and magnitude-cut effect)\\
& $\pm$2 & Modeling error (scaling error to 2MASS counts at brighter magnitudes,\\
&& and uncertainty of galaxy counts at fainter magnitudes)\\
DGL & 20\% ($\pm$4 at NEP) & Uncertainty of DGL/100 $\mu$m ratio \citep{2015ApJ...806...69A}\\
&&\\
EBL field mean & $\pm$2 & Statistical error and field-to-field variation\\
(ZL model) & $+$12/$-$11 & Systematic uncertainty including absolute calibration,\\
&& airglow modeling error, and ISL and DGL modeling errors\\
&&\\
EBL field mean & $\pm$3 & Statistical error, field-to-field variation, and scaling error to IGL at 0.8--1.0 $\mu$m\\
(Minimum) & $+$4/$-$1 & Systematic uncertainty including ZL template error,\\
&& spectral calibration error, and airglow modeling error\\
\enddata
\tablenotetext{a}{in nWm$^{-2}$sr$^{-1}$ at 1.25 $\mu$m, or percent}
\tablenotetext{b}{Negligible when a ZL template from a certain flight is used for same flight data}
\end{deluxetable}

\section{Discussions}
If the EBL excess arises mainly from high-redshift objects, it obviously conflicts with the gamma-ray transparency of the intergalactic medium (IGM) corresponding to the number of EBL photons for the electron-positron pair creation, as discussed by many authors on previous EBL results \citep{2006Natur.440.1018A,2009A&A...498...25R,2013A&A...550A...4H}.  Therefore, the EBL excess would be attributed to an unaccounted component in the local universe where the gamma-ray absorption length in the line of sight is relatively short, unless new physical mechanisms which reduce the intergalactic gamma-ray absorption \citep{2008PhLB..659..847D,2010PhRvL.104n1102E} are particularly required.  In fact, an overall EBL spectrum, inferred from our nominal result and previous near-infrared data, is simply analogous to that of the redder galaxies at low redshifts; it exhibits a broad peak around $\sim$1.5 $\mu$m and the Rayleigh-Jeans tail at longer wavelengths.  While there is no room to account a large number of missing galaxies in the local universe, unresolvable diffuse light of old stellar populations in galaxy halo and/or intergalactic medium may be source of the EBL excess having redder color.  However, it should be noted that faint galaxy wings of resolved galaxies have little contribution to the EBL \citep{2010ApJS..186...10A,2015MNRAS.449.1291D}.

Our minimum EBL result exhibits a significant excess over the IGL level, but it is marginally consistent with the EBL limit obtained from the gamma-ray observations, and such a relatively small discrepancy could be due to an unaccounted source at high redshift.  The gamma-ray limits on the EBL assume a specific spectral shape for the EBL.  For a different spectral shape and a different EBL photon energy density as a function of redshift, the gamma-ray limits may be higher in a limited wavelength range as to accord with our minimum EBL result.

Intrahalo light (IHL), produced by tidally stripped stars in low-redshift galaxies, contributes significantly to the near-infrared background fluctuations \citep{2012Natur.490..514C,2015NatCo...6E7945M}, indicated by the excess clustering power on arcminutes scales \citep{2005Natur.438...45K,2007ApJ...657..669T,2011ApJ...742..124M,2014Sci...346..732Z}.  The IHL contribution to the EBL excess in the LRS bands, from the model fitting to the fluctuations measured with the {\it CIBER} Imager instrument, is 7.0 and 11.4 nWm$^{-2}$sr$^{-1}$ at 1.1 and 1.6 $\mu$m, respectively, based on a relative EBL fluctuation amplitude of $\delta$$I$/$I$$\sim$20\%, which is model dependent and could vary from 10-30\% \citep{2014Sci...346..732Z}.  Our measured EBL excess over the IGL is 25.6$\pm{12.9}$ and 29.7$^{+15.9}_{-9.9}$ nWm$^{-2}$sr$^{-1}$ at 1.1 and 1.6 $\mu$m for the nominal EBL, respectively, and is 4.4$^{+3.6}_{-3.5}$ and 15.6$^{+10.7}_{-4.1}$ nWm$^{-2}$sr$^{-1}$ at 1.1 and 1.6 $\mu$m for the minimum EBL, respectively.  Therefore, IHL may account for some of the EBL excess.  More precise measurements are needed to compare the color between the intensity and fluctuations of the EBL.

First-light galaxies during the epoch of reionization (EoR), including PopII/PopIII stars \citep{2004MNRAS.351L..71C,2006ApJ...646..703F,2013ApJ...768..197I} and direct collapse blackholes \citep{2013MNRAS.433..1556Y}, predict a distinctive spectral shape of the EBL excess, which peaks at $\sim$1.5 $\mu$m analogous to the Lyman break.  Such objects with $\delta$$I$/$I$$\sim$5\% can contribute effectively more to the mean EBL intensity than the fluctuations.  However, an EBL intensity from the EoR as high as $\sim$10 nWm$^{-2}$sr$^{-1}$ has difficulties due to the over-enrichment of the intergalactic medium by metals and an overproduction of the X-ray background \citep{2005MNRAS.359L..37M}.  For a currently favored star-formation rate density at $z >$ 7 \citep{2012ApJ...754...83B}, the high-$z$ contribution to the EBL is at the level of 0.5--1 nWm$^{-2}$sr$^{-1}$.

Dark stars, hypothetical objects powered by annihilation of accreted weakly interacting massive particles (WIMPs) in the early universe, have been proposed to produce an EBL peaking at $\sim$2 $\mu$m \citep{2012ApJ...745..166M}.  Certain dark star models can be made consistent with the observed EBL excess (because their overall normalization is not constrained), but would be in conflict with the gamma-ray limits.

Even though the minimum EBL result is obtained by conservative subtraction of the ZL foreground, a spatially isotropic ZL component could still be present \citep{1998ApJ...508...25H}.  However, if this is the case, the dust grains must have an extraordinary spectral reflectance \citep{1995Icar..115..199M}.  Because the EBL residuals from our data do not have a ZL spectrum, an isotropic ZL component cannot explain our results.  A definitive conclusion on the origin of the near-infrared EBL would be most easily achieved by an instrument in deep space \citep{matsuura02,matsuura13} where the ZL foreground is absent \citep{1974JGR....79.3665W,1974JGR....79.3671H,2011ApJ...736..119M}\footnote{The background observations with {\it Pioneer10/11} were not accurate enough to significantly detect the optical EBL.}, or by precise multi-band fluctuation measurements \citep{2014SPIE.9143E..3NL,2016SPIE.9904E..4JS}.

\section{Summary}
We achieved the first direct measurement of the diffuse background spectrum at 0.8--1.7 $\mu$m with the {\it CIBER} experiment.  The EBL spectrum is derived using model-based foreground assessments of the ISL, DGL and ZL components.  The residual absolute surface brightness, 42.7$^{+11.9}_{-10.6}$ nWm$^{-2}$sr$^{-1}$ at 1.4 $\mu$m, exceeds both the IGL amplitude and upper limits from the gamma-ray transparency of IGM.  These measurements are consistent with previous results from the satellite observations at $>$1.25 $\mu$m within the measurement errors, if we compare to those analyses assuming the Kelsall ZL model.

If we assume the Wright ZL model, the derived EBL spectrum is unphysical, becoming less than the IGL at $\lambda<$1.3 $\mu$m and negative at $\lambda<$1 $\mu$m.  For both ZL models, the derived EBL is redder than the ZL spectrum at $\lambda<$1.4 $\mu$m.  The variation we observe between fields more closely matches the Kelsall ZL model than the Wright ZL model.

The minimum EBL levels at $>$1.25 $\mu$m are estimated using a model-independent method of the ZL subtraction by scaling the ZL template so that the average surface brightness of the EBL for 0.8 $< \lambda <$ 1 $\mu$m equals the IGL.  The minimum EBL brightness, \textit{e.g.}, 28.7$^{+5.1}_{-3.3}$ nWm$^{-2}$sr$^{-1}$ at 1.4 $\mu$m, significantly exceeds the IGL level, and its spectrum exhibits a red color quite different from the blue continuum of the ZL.

These results could be explained by a new foreground component, such as color variations in the ZL, or by a new EBL component due to unaccounted emission from galaxies.  Further observational studies in wider wavelength range by future space missions are needed to improve the accuracy of the EBL measurements, due to the uncertainties with the ZL subtraction.  For the near term, we plan to extend the range of these absolute spectral measurements to include optical wavelengths.

\acknowledgments
We acknowledge the support of Grant-in-Aid from Japan Society for the Promotion of Science (KAKENHI 20.34, 18204018, 19540250, 21340047, 21111004, 24111717, 26800112 and 15H05744) and NASA APRA research grant which funded US contributions to the CIBER instrument (NNX07AI54G, NNG05WC18G, NNX07AG43G, NNX07AJ24G, and NNX10AE12G).  Initial support was provided by an award to J.B. from the Jet Propulsion Laboratory's Director's Research and Development Fund.  We thank the dedicated efforts of the sounding rocket staff at NASA Wallops Flight Facility and White Sands Missile Range, and also thank Dr. Allan Smith, Dr. Keith Lykke, and Dr. Steven Brown (NIST) for the laboratory calibration of LRS.  A.C. acknowledges support from an NSF CAREER award AST-0645427 and NSF AST-1313319.  H.M.L. was supported by
NRF grant No.2012R1A4A1028713.  S.M. thanks to Dr. Fumihiko Usui (Kobe University, CPS) and Dr. Takafumi Ootsubo (The University of Tokyo) for their help in coordinate calculations and for our discussions on the ZL models.  This publication makes use of data products from the {\it 2MASS}, which is a joint project of the University of Massachusetts and the Infrared Processing and Analysis Center/California Institute of Technology, funded by the National Aeronautics and Space Administration and the National Science Foundation.

\appendix

\section{Airglow subtraction method}\label{airglow}
\subsection{A Simple Airglow Model with Exponential Functions} 
\begin{figure}
\epsscale{0.6}
\plotone{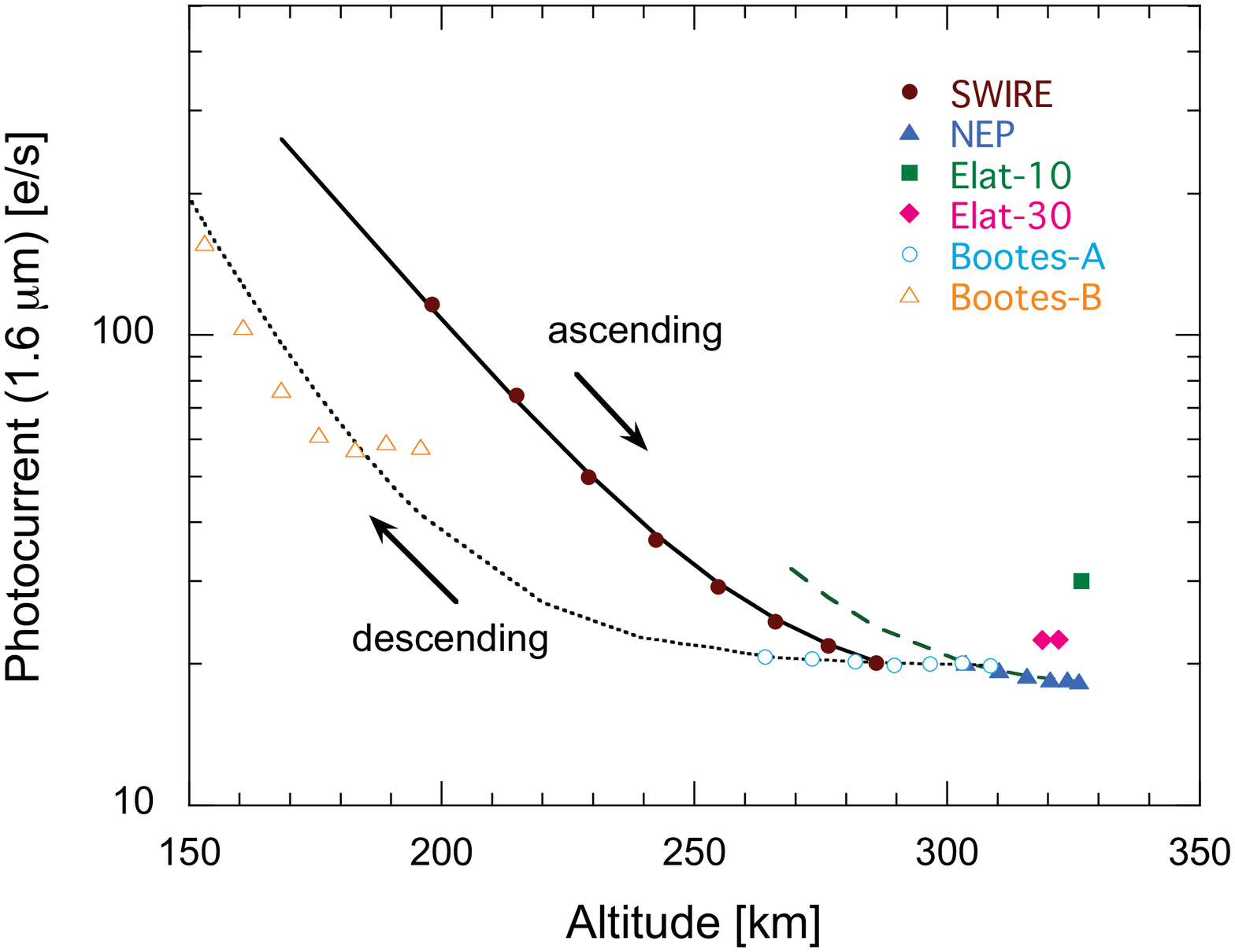}
\caption{Altitude dependence of the airglow seen in the second flight.  The observed photocurrent signals at the peak wavelength of the airglow spectrum, 1.6 $\mu$m, are plotted against the altitude.  The solid, dashed and dotted curves indicate the result of the exponential fitting to the data at SWIRE, NEP and Bootes-A/B, respectively, with $\tau$=20 s and $H$=22 km.\label{figA1}}
\end{figure}

As shown in Figure~\ref{figA1}, the airglow brightness at a given wavelength can be approximately fit by exponential functions modeled as the sum of time- and altitude-dependent components:

\begin{equation}
I_{obs}=I_{sky}+A_{t}e^{-t/\tau}+A_{h}e^{-h/H}
\end{equation}

where the time constant is $\tau$, the scale height is $H$, $t$ is the elapsed time after launch, $h$ is the payload altitude, and$I_{obs}$ and $I_{sky}$ are the observed raw and astronomical brightnesses, respectively.  The amplitude parameters, $A_t$ and $A_h$, give the brightness of the time-dependent and altitude-dependent airglow components at a given wavelength at $t=0$ and $h=0$, respectively.  The time- and altitude-dependent components are indistinguishable during the ascent phase.  As a result, we fit the altitude-dependent component to the last science field (Bootes-A of the second flight) of the descent phase when the time-dependent component is negligible.  We then fit the time-dependent component to the first science field (SWIRE of the second flight) after subtracting an altitude-dependent component.  The model curves shown in Figure~\ref{figA1} are obtained by fitting to the data with $\tau$=20 s and $H$=22 km, which are fixed for all wavelengths to the values determined at the peak wavelength, i.e., 1.6 $\mu$m.

Using this result, we can successfully subtract the airglow component to negligible levels compared to the sky brightness.  Airglow suppression is not necessary for fields near the apogee of the flight.  The SWIRE data showed residual time dependence early in the observations even after the airglow model was subtracted, but the final quarter of the SWIRE data appears stable.  The altitude dependence at Bootes-B showed anomalous behavior, which may be due to complex layering in atmospheric OH at lower altitudes, and we did not use these data for astrophysical analysis.  For the second flight the final high-quality data set consists of five sky spectra at SWIRE, NEP, Elat10, Elat30, and Bootes-A in 0.8 $< \lambda <$ 1.7 $\mu$m.

\subsection{Existence of slow decay airglow component}\label{slowag}
\begin{figure}
\epsscale{0.6}
\plotone{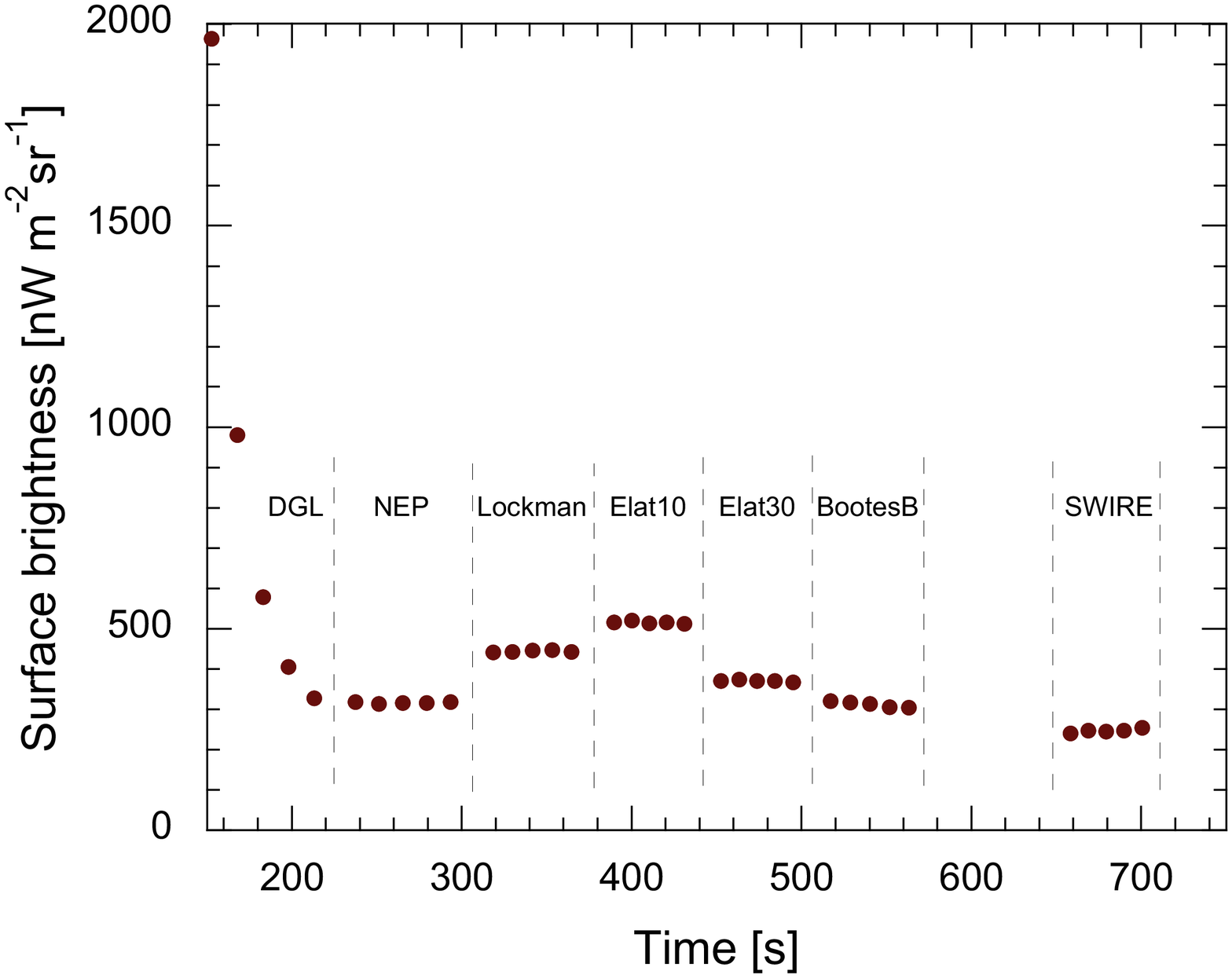}
\caption{Slow-decay airglow component.  The raw sky brightness at 1.6 $\mu$m before airglow subtraction is plotted as a function of the elapsed time after the launch.  The brightness variations are mainly due to difference in the sky brightness of different fields, except at the beginning and end of the flight, when the airglow is strong compared with the sky brightness.\label{figA2}}
\end{figure}

The above airglow model works only for a fast decay component with a time constant shorter than the integration time of a field.  Although we can determine the decay time constant of the time-dependent airglow component from the second flight data alone, there may exist an unaccounted component with a time constant exceeding the integration time, \textit{i.e.}, $\tau >$100 s.  We observe such a slow airglow component in the fourth flight data, which is sufficiently long for monitoring very slow signal variations.  The time variation of the fourth flight at 1.6 $\mu$m (corresponding to the maximum brightness of the airglow) is shown in Figure~\ref{figA2}.  The fast decay airglow component is prominent early in the observations.  Owing to the high altitude of the fourth flight, an altitude-dependent airglow component is not observed during the flight, so these data are used to constrain slow signal variations.

Since the sky brightness varies from field to field, the slowly decaying airglow component is not easily observed in the raw brightness data shown in Figure~\ref{figA2}.  In order to extract the time variation of the airglow from the sky brightness, we use the ratio of the sky brightness at 1.6 $\mu$m, the maximum airglow wavelength, to the sky brightness at 0.9 $\mu$m, the minimum airglow wavelength (see Figure~\ref{fig3}).  The 1.6-$\mu$m to 0.9-$\mu$m brightness ratios are nearly constant when the ZL dominates the total sky brightness.  Even assuming an EBL twice as bright as the known IGL brightness, the color variations of the total sky brightness due to the differing contributions of ISL, DGL and EBL to the total sky brightness are estimated to be less than 2\% rms, corresponding to a sky brightness of about 6 nWm$^{-2}$sr$^{-1}$ at 1.6 $\mu$m.  Therefore, any time variation of the 1.6 $\mu$m to 0.9 $\mu$m ratio exceeding the 2\% level must be attributed to non-astrophysical emission.

\begin{figure}
\epsscale{0.6}
\plotone{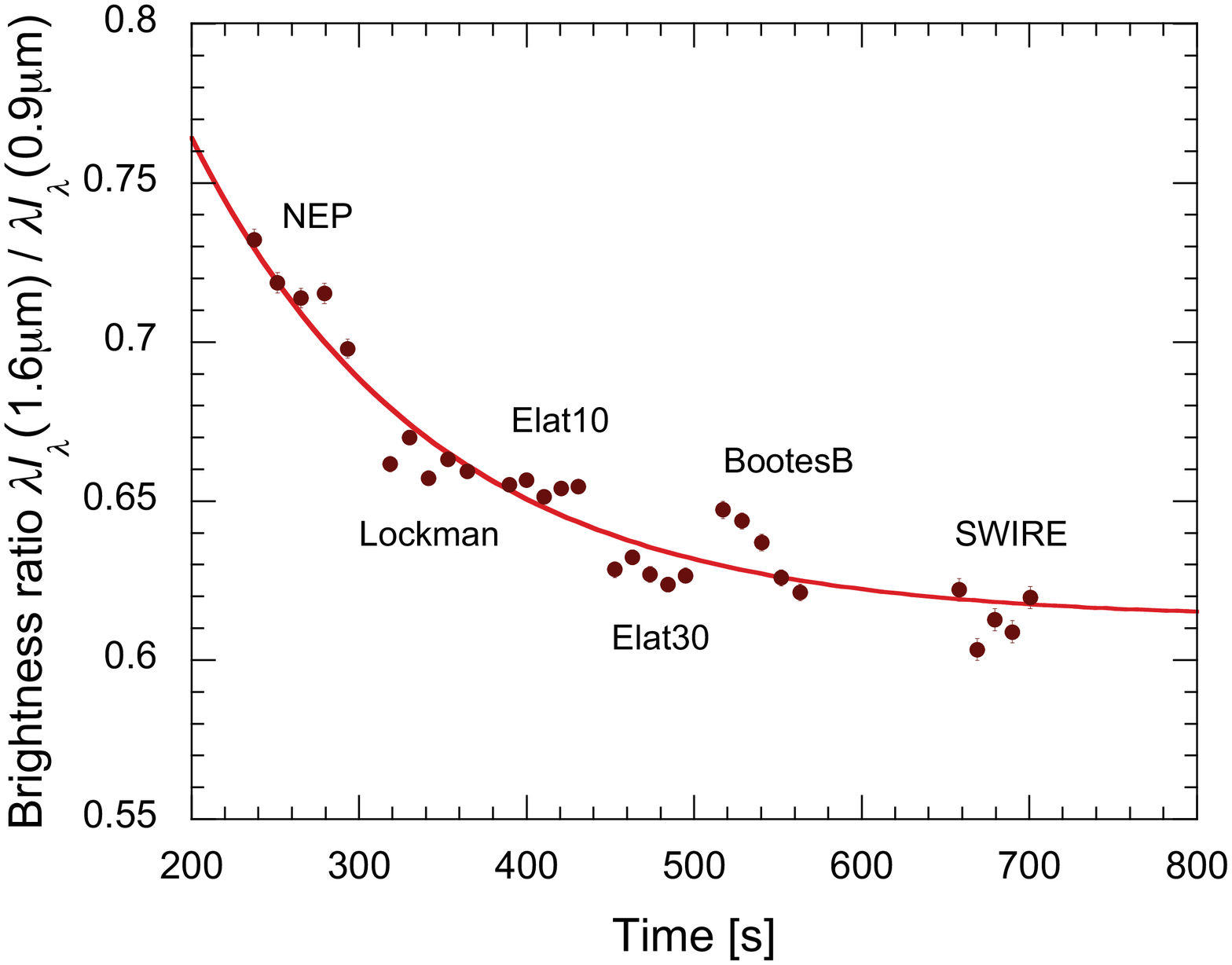}
\caption{The 1.6-$\mu$m to 0.9-$\mu$m brightness ratio during the fourth flight.  The ratio is plotted as a function of the elapsed time after launch.  A trend of monotonic decrease with a long time constant is clearly seen.  The data can be fit by a single exponential function (solid line).  The fast-decay component is pre-subtracted before fitting.\label{figA3}}
\end{figure}

\begin{figure}
\epsscale{0.6}
\plotone{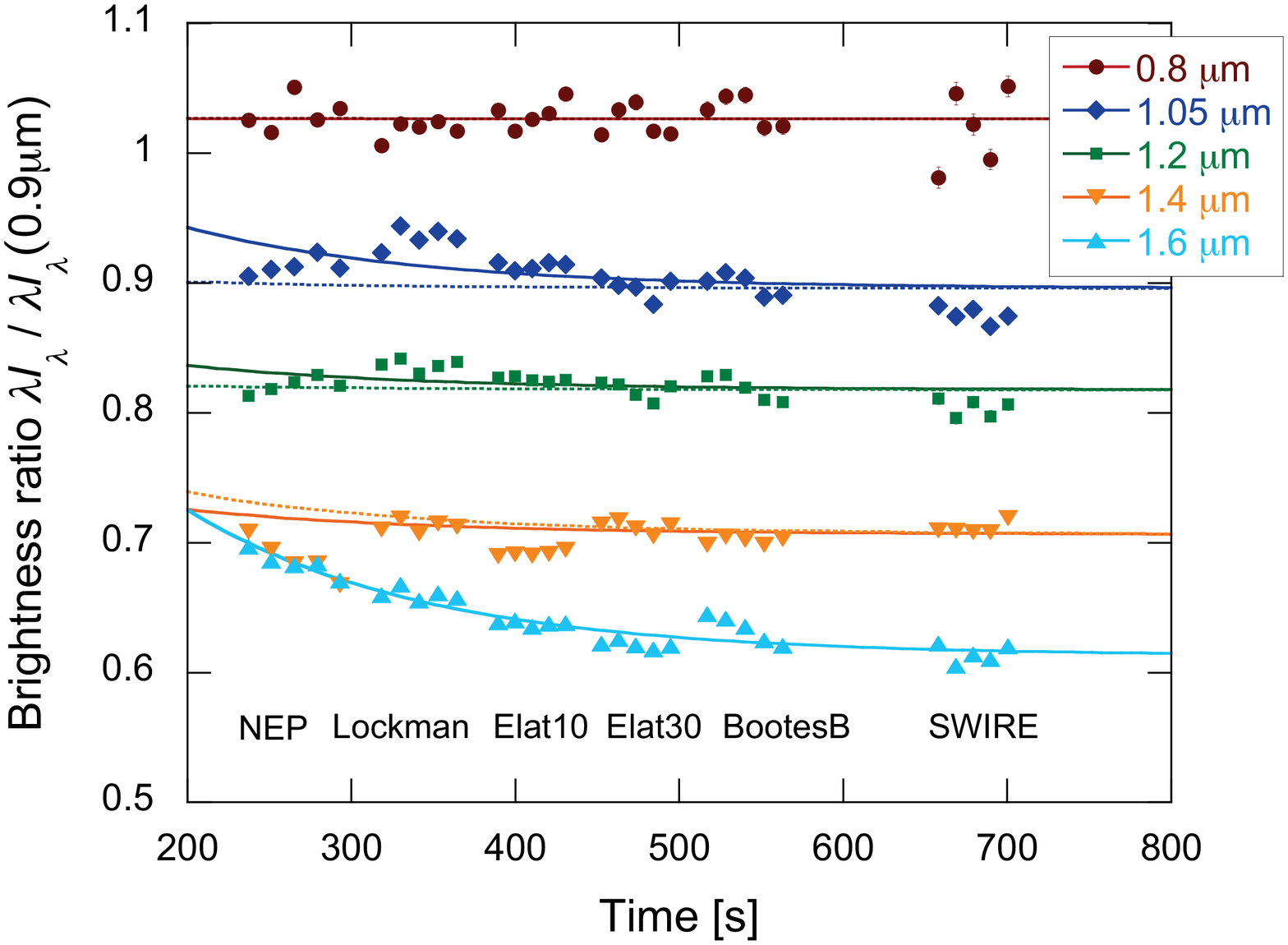}
\caption{The brightness ratios at various wavelengths.  The data at 1.6 $\mu$m are same as shown in Figure~\ref{figA3}.  The solid lines are exponential functions fit to the data at each wavelength with a fixed time constant of the slow-decay airglow component obtained from the data at 1.6 $\mu$m.  The dotted lines are the model curves with a spectral shape of the fast decay airglow.  The model fits the data well, except for the 1.05-$\mu$m data.\label{figA4}}
\end{figure}

\begin{figure}
\epsscale{0.6}
\plotone{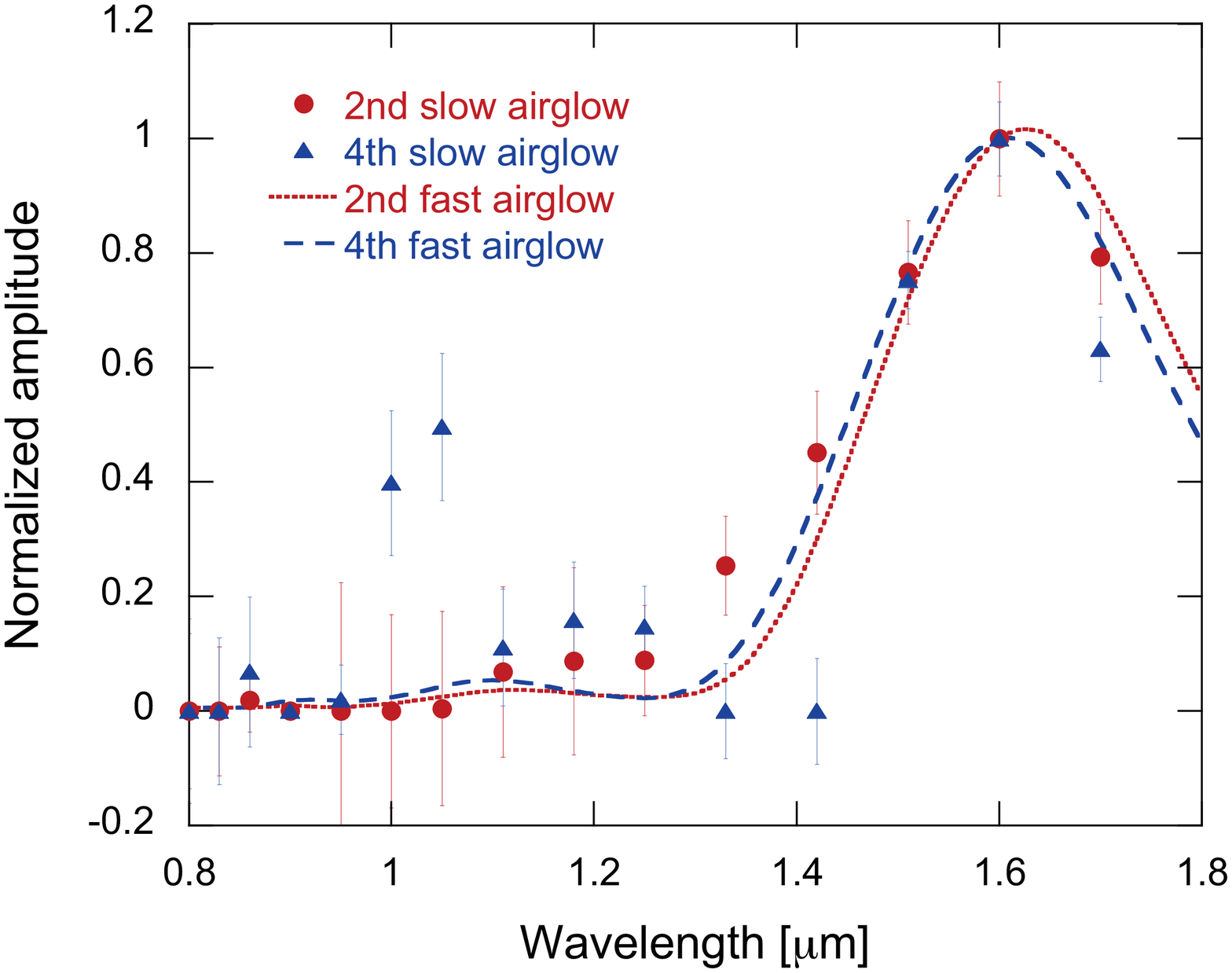}
\caption{Spectral shape of the slow decay component of the airglow.  The signal amplitudes for the fourth flight indicated by the filled triangles are obtained from the exponential fitting to the data as shown in Figure~\ref{figA4}.  The filled circles denote the slow decay airglow component seen in the second flight derived by fitting the airglow model to the second flight data.  The dotted and dashed lines are normalized spectra of the fast decay airglow component seen at the time early in the second and fourth flight, respectively.  The slow decay airglow spectra showing peak at 1.6 $\mu$m are similar to the fast decay airglow spectrum, though an unidentified feature at 1.0--1.05 $\mu$m is seen for the fourth flight data. Thus, the fast decay airglow spectrum is used as a spectral model template for the slow decay airglow component.\label{figA5}}
\end{figure}

As shown in Figure~\ref{figA3}, the measured 1.6 $\mu$m to 0.9 $\mu$m brightness ratio for the fourth flight exhibits a slow component (in addition to the fast component seen early in the observation), and the ratio of the SWIRE field reaches an asymptotic brightness similar to the predicted value for astronomical sources of 0.62.  The slow-decay component can be fit by an exponential function with a time constant of $\tau_{slow}$=144$\pm$19 s, while the fast-decay component can be fit with $\tau_{fast}$=17.5$\pm$0.2 s.  Figure~\ref{figA4} shows the relative sky brightness at various wavelengths normalized by the sky brightness at 0.9 $\mu$m.  These data are fit by exponential functions with a time constant fixed to $\tau_{slow}$=144 s, setting the amplitudes free.  The spectral shape of the determined amplitude parameters with a prominent broad peak around 1.6 $\mu$m is similar to that of the fast decay airglow component, except for a feature around 1 $\mu$m, as shown in Figure~\ref{figA5}.  The 1 $\mu$m feature is not instrumental effect but real signal, because it is seen in the FOV area of the multiple slits but not in the masked area of the detector array.

\subsection{Application of the Airglow Model}\label{agmodel}
\begin{figure}
\epsscale{0.6}
\plotone{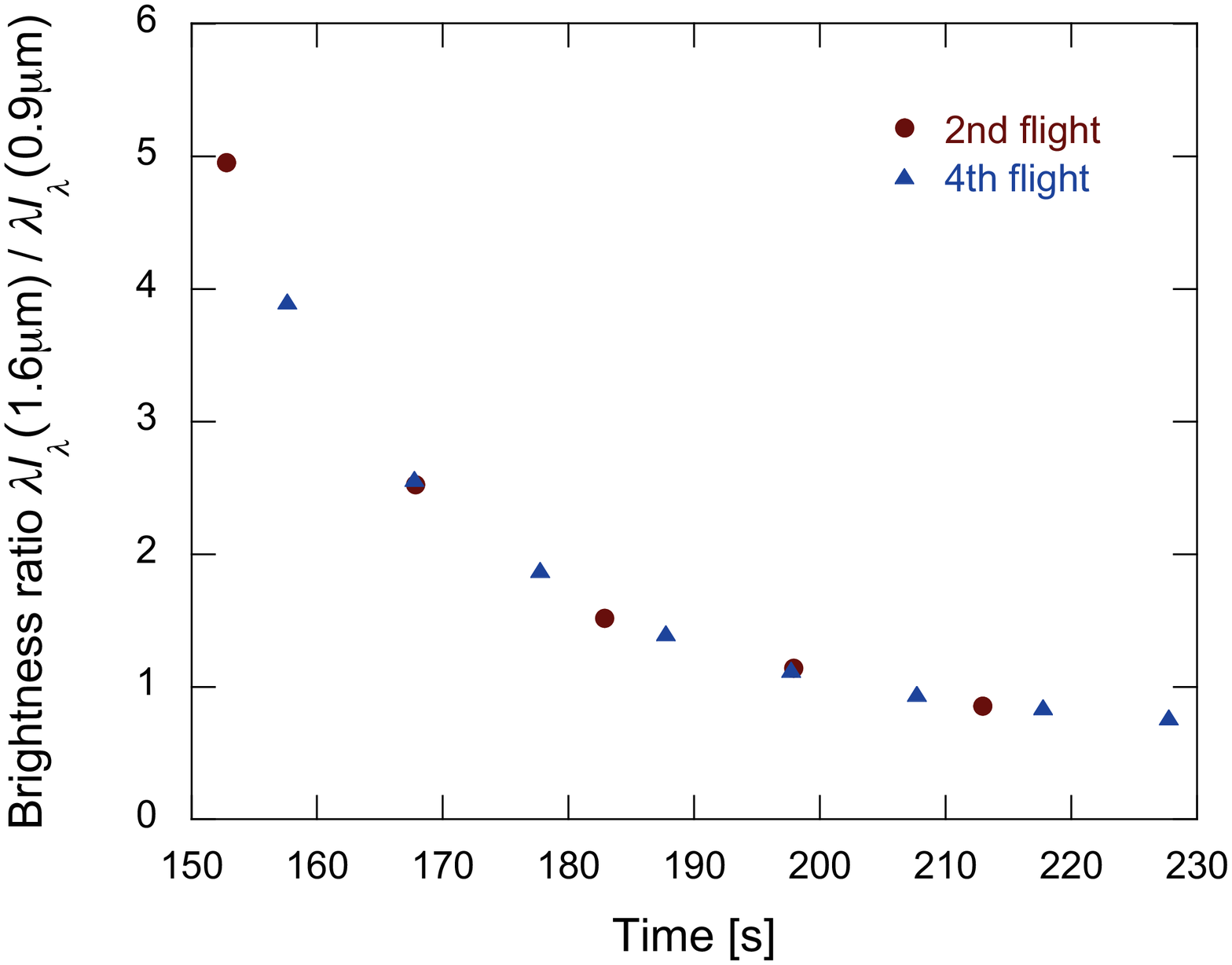}
\caption{Comparison of the airglow decay trend between two flights.  In early observations, the time dependence of the 1.6-$\mu$m to 0.9-$\mu$m brightness ratio of the second (circles) and fourth (triangles) flight agrees with each other.  The time offsets are shifted to adjust the amplitudes, as the emission is earlier in the fourth flight.  This common behavior motivates us to subtract the slow-decay component from not only the fourth flight data but also the second flight data.\label{figA6}}
\end{figure}

This slow decay airglow component may also be present in the second flight data, but unobservable due to the short time scale of the flight.  The airglow signal has a large derivative early in the observation when the time-dependent airglow dominates, so are best compared using the 1.6-$\mu$m to 0.9-$\mu$m brightness ratio as shown in Figure~\ref{figA6}.  After adjusting for a time offset related to the time of opening the shutter door since ignition of the rocket, the shapes of the decays in the two flights are very similar.  This result suggests that the second flight may be contaminated by the same airglow seen in the fourth flight.

\begin{figure}
\epsscale{0.6}
\plotone{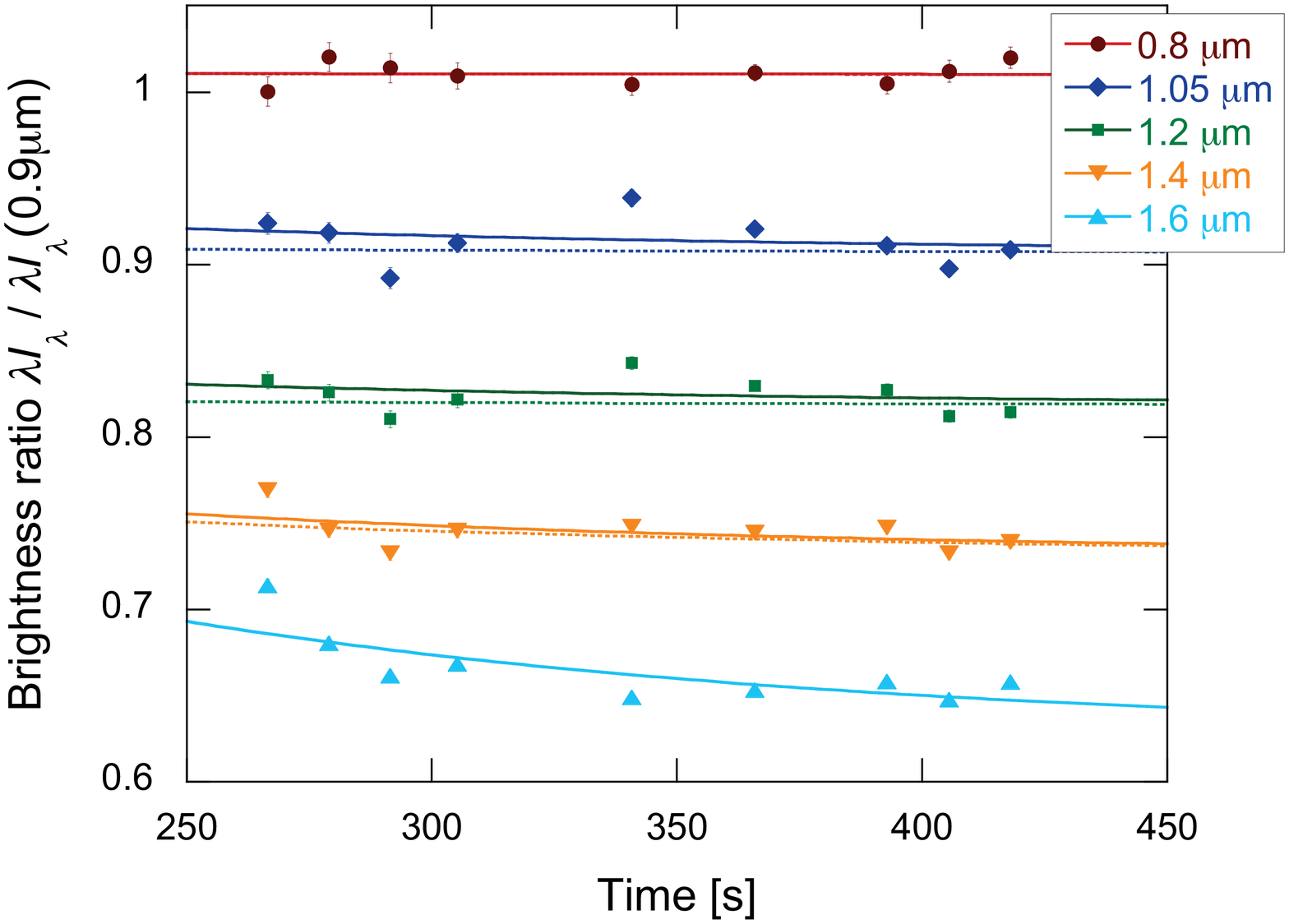}
\caption{The airglow model applied to the second flight data.  The brightness ratios at various wavelengths are plotted as functions of the elapsed time after launch.  The fast decay airglow is pre-subtracted. The solid and dotted lines are same as in Figure~\ref{figA4}; exponential fits with a fixed time constant from the fourth flight with free amplitudes and fixed amplitudes with the template spectrum as shown in Figure~\ref{figA5}, respectively.  Even late in the observation ($\sim$400 s), the modeled airglow intensity is still decreasing.\label{figA7}}
\end{figure}

The fast decay airglow spectra in the two flights are similar, as shown in Figure~\ref{figA5}.  Because of this, we fit the slow decay airglow spectrum in the second flight using the fixed time constant measured in the fourth flight, but with free amplitude parameters at each wavelength.  The derived spectrum of the amplitude parameters is consistent with the fast decay airglow spectrum, as shown in Figures~\ref{figA5} and \ref{figA7}.  For simplicity we assume that the slow decay airglow component has same spectral shape as the fast decay airglow component in both flights.  From this modeling we can calculate the airglow intensity at a given wavelength and time using our knowledge of the 1.6 $\mu$m to 0.9 $\mu$m brightness ratio and the spectral template.

Our nominal result of the sky brightness measurement is obtained by subtracting the modeled airglow brightness.  The fitting error of the exponential function is accounted as statistical error of the sky brightness measurement.  For the second flight the full brightness difference with and without subtraction of the slow-decay component is conservatively taken as the systematic error associated with airglow subtraction.


\begin{thebibliography}{}

\bibitem[Aharonian et al.(2006)]{2006Natur.440.1018A} Aharonian, F., Akhperjanian, A.G., Bazer-Bachi, A. R., et al.\ 2006, {\it Nature}, 440, 1018
\bibitem[Arai et al.(2015)]{2015ApJ...806...69A} Arai, T., Matsuura, S., Bock, J., et al.\ 2015, \apj, 806, 69
\bibitem[Arendt et al.(2010)]{2010ApJS..186...10A} Arendt, R. G., Kashlinsky, A., Moseley, S.H., et al.\ 2010, \apjs, 186, 10
\bibitem[Bouwens et al.(2012)]{2012ApJ...754...83B} Bauwens, R. J., Zheng, W., Moustakas, L., et al.\ 2012, \apj, 754, 83
\bibitem[Bernstein(2007)]{2007ApJ...666..663B} Bernstein, R. A.\ 2007, \apj, 666, 663
\bibitem[Bock et al.(2013)]{2013ApJS..207...32B} Bock, J., Sullivan, I., Arai, T., et al.\ 2013, \apjs, 207, 32
\bibitem[Brandt \& Draine(2012)]{2012ApJ...744..129B} Brandt, T. D. \& Draine B. T.\ 2012, \apj, 744, 129
\bibitem[Cambr\`esy et al.(2001)]{2001ApJ...555..563C} Cambr\`esy, L., Reach, W. T., Beichman, C. A., et al.\ 2001, \apj, 555, 563
\bibitem[Cooray \& Yoshida(2004)]{2004MNRAS.351L..71C} Cooray, A., \& Yoshida, N.\ 2004, \mnras, 351, L71
\bibitem[Cooray et al.(2012)]{2012Natur.490..514C} Cooray, A., Smidt, J., de Bernardis, F., et al.\ 2012, {\it Nature}, 490, 514
\bibitem[de Angelis et al.(2008)]{2008PhLB..659..847D} de Angelis, A., Mansutti, O., \& Roncadelli, M.\ 2008, Phys. Lett. B, 659, 847
\bibitem[Dom\'inguez et al.(2011)]{2011MNRAS.410.2556D} Dom\'inguez, A., Primack, J. R., Rosario, D. J., et al.\ 2011, \mnras, 410, 2556
\bibitem[Donnerstein(2015)]{2015MNRAS.449.1291D} Donnerstein, R.L.\ 2015, \mnras, 449, 1291
\bibitem[Dwek et al.(2005)]{2005ApJ...635..784D} Dwek, E., Arendt, R. G., \& Krennrich, F.\ 2005, \apj, 635, 784
\bibitem[Essey et al.(2010)]{2010PhRvL.104n1102E} Essey, W., Kalashev, O. E., Kusenko, A., \& Beacom, J. F.\ 2010, \prl, 104, 141102
\bibitem[Fernandez \& Komatsu(2006)]{2006ApJ...646..703F} Fernandez, E. \& Komatsu, E.\ 2006, \apj, 646, 703
\bibitem[Girardi et al.(2005)]{2005A&A...436..895G} Girardi, L., Demarco, R., Rosati, P., \& da Costa, L.\ 2005, \aap, 436, 895
\bibitem[Gonzalez et al.(2010)]{2010AAS...21641513G} Gonzalez, A. H., Brodwin, M., Brown, M. J. I., et. al.\ 2010, {\it AAS}, 216, 415.13
\bibitem[Hanner et al.(1974)]{1974JGR....79.3671H} Hanner, M. S., Weinberg, J. L., DeShields II, L. M., et al.\ 1974, \jgr, 79, 3671
\bibitem[Hauser et al.(1998)]{1998ApJ...508...25H} Hauser, M., Arendt, R. G., Kelsall, T., et al.\ 1998, \apj, 508, 25
\bibitem[Hauser \& Dwek(2001)]{2001ARA&A..39..249H} Hauser, M.G. \& Dwek, E.\ 2001, \araa, 39, 249
\bibitem[H.E.S.S. Collaboration(2013)]{2013A&A...550A...4H} H.E.S.S. Collaboration\ 2013, \aap, 550, A4
\bibitem[Inoue et al.(2013)]{2013ApJ...768..197I} Inoue, Y., Inoue, S., Kobayashi, M. A. R., et al.\ 2013, \apj, 768, 197
\bibitem[Kashlinsky et al.(2005)]{2005Natur.438...45K} Kashlinsky, A., Arendt, R. G., Mather, J., \& Moseley, S. H.\ 2005, {\it Nature}, 438, 45
\bibitem[Keenan et al.(2010)]{2010ApJ...723...40K} Keenan, R. C., Barger, A. J., Cowie, L. L., \& Wang, W.-H.\ 2010, \apj, 723, 40
\bibitem[Kelsall et al.(1998)]{1998ApJ...508...44K} Kelsall, T., Weiland, J. L., Franz, B. A., Reach, W. T., et al.\ 1998, \apj, 508, 44
\bibitem[Kim et al.(2017)]{2017AJ....153...84K} Kim, M. G., Lee, H. M., Jeong, W.-S., et al.\ 2017, \aj, 153, 84
\bibitem[Korngut et al.(2013)]{2013ApJS..207...34K} Korngut P. M., Renbarger, T., Arai, T., et al.\ 2013, \apjs, 207, 34
\bibitem[Lanz et al.(2014)]{2014SPIE.9143E..3NL} Lanz, A., Arai, T., Battle, J., et al.\ 2014, {\it Proc. SPIE}, 9143, 91433N
\bibitem[Levenson et al.(2007)]{2007ApJ...666...34L} Levenson, L. R., Wright, E. L., \& Johnson, B. D.\ 2007, \apj, 666, 34
\bibitem[Madau \& Pozzeti(2000)]{2000MNRAS.312L...9M} Madau, P. \& Pozzeti, L.\ 2000, \mnras, 312, L9
\bibitem[Madau \& Silk(2005)]{2005MNRAS.359L..37M} Madau, P. \& Silk, J.\ 2005, \mnras, 359, L37
\bibitem[Matsumoto et al.(2005)]{2005ApJ...626...31M} Matsumoto, T., Matsuura, S., Murakami, H., et al.\ 2005, \apj, 628, 31
\bibitem[Matsumoto et al.(2011)]{2011ApJ...742..124M} Matsumoto, T., Seo, H. J., Jeong, W.-S., et al.\ 2011, \apj, 742, 124
\bibitem[Matsumoto et al.(2015)]{2015ApJ...807...57M} Matsumoto, T., Kim, M. G., Pyo, J., \& Tsumura, K.\ 2015, \apj, 807, 57
\bibitem[Matsuoka et al.(2011)]{2011ApJ...736..119M} Matsuoka Y., Ienaka, N., Kawara, K., \& Oyabu, S.\ 2011, \apj, 736, 119
\bibitem[Matsuura et al.(1995)]{1995Icar..115..199M} Matsuura, S., Matsumoto, T., Matsuhara, H., \& Noda, M.\ 1995, Icarus, 115, 199
\bibitem[Matsuura(2002)]{matsuura02} Matsuura, S.\ 2002, {\it Proc. Far-IR, Sub-mm \& MM Detector Technology Workshop}, ed. J. Wolf, J. Farhoomand, \& C. R. McCreight, NASA/CP-211408, i-04
\bibitem[Matsuura et al.(2013)]{matsuura13} Matsuura, S., Yano, H., Yonetoku, D., et al.\ 2013, {\it Trans. JSASS, Aerospace Tech. Japan}, 29, 2013-r-09
\bibitem[Mattila et al.(2011)]{mattila11} Mattila K., Lehtinen, K., V\"ais\"anen, P., et al.\ 2011, {\it Proc. IAU Symposium}, 284, 429
\bibitem[Maurer et al.(2012)]{2012ApJ...745..166M} Maurer, A., Raue, M., Kneiske, T., et al.\ 2012, \apj, 745, 166
\bibitem[Mitchell-Wynne et al.(2015)]{2015NatCo...6E7945M} Mitchell-Wynne, K., Cooray, A., Gong, Y., et al.\ 2015, Nature communications, 6, 7945
\bibitem[Oi et al.(2014)]{2014A&A...566A..60O} Oi, N., Matsuhara, H., Murata, K., et al.\ 2014, \aap, 566, A60
\bibitem[Raue et al.(2009)]{2009A&A...498...25R} Raue, M., Kneiske, T., \& Mazin, D.\ 2009, \aap, 498, 25
\bibitem[Sano et al.(2015)]{2015ApJ...811...77S} Sano, K., Kawara, K., Matsuura, S., et al.\ 2015, \apj, 811, 77
\bibitem[Sano et al.(2016)]{2016ApJ...821L..11S} Sano, K., Matsuura, S., Tsumura, K., et al.\ 2016, \apjl, 818, 72
\bibitem[Santos et al.(2002)]{2002MNRAS.336.1082S} Santos, M. R., Bromm, V., \& Kamionkowski, M.\ 2002, \mnras, 336, 1082
\bibitem[Salvaterra et al.(2003)]{2003MNRAS.339..973S} Salvaterra, R., \& Ferrara, A.\ 2003, \mnras, 339, 973
\bibitem[Shirahata et al.(2016)]{2016SPIE.9904E..4JS} Shirahata, M., Arai, T., Battle, J., et al.\ 2016, {\it Proc. SPIE}, 9904, 99044J
\bibitem[Skrutskie et al.(2006)]{2006AJ....131.1163S} Skrutskie, M. F., Cutri, R. M., Stiening, R., et al.\ 2006, \aj, 131, 1163
\bibitem[Thompson et al.(2007)]{2007ApJ...657..669T} Thompson, R. I., Eisenstein, D., Fan, X., et al.\ 2007, \apj, 657, 669
\bibitem[Tsumura et al.(2010)]{2010ApJ...719..394T} Tsumura, K., Battle, J., Bock, J., et al.\ 2010, \apj, 719, 394
\bibitem[Tsumura et al.(2013a)]{2013ApJS..207...33T} Tsumura, K., Arai, T., Battle, J., et al.\ 2013, \apjs, 207, 33
\bibitem[Tsumura et al.(2013b)]{2013PASJ...65..120T} Tsumura, K., Matsumoto, T., Matsuura, S., et al.\ 2013, \pasj, 65, 120
\bibitem[Tsumura et al.(2013c)]{2013PASJ...65..121T} Tsumura, K., Matsumoto, T., Matsuura, S., et al.\ 2013, \pasj, 65, 121
\bibitem[Weinberg et al.(1974)]{1974JGR....79.3665W} Weinberg, J. L., Hanner, M. S., Beeson, D. E., et al.\ 1974, \jgr, 79, 3665
\bibitem[Wright(1998)]{1998ApJ...496....1W} Wright, E. L.\ 1998, \apj, 496, 1
\bibitem[Wright \& Reese(2000)]{2000ApJ...545...43W} Wright, E. L. \& Reese, E. D.\ 2001, \apj, 545, 43
\bibitem[Wright(2001)]{2001IAUS..204..157W} Wright, E. L.\ 2001, {\it Proc. IAU Symposium}, 204, 157
\bibitem[Yue et al.(2013a)]{2013MNRAS.431..383Y} Yue, B., Ferrara, A., Salvaterra, R. \& Chen, X.\ 2013, \mnras, 431, 383
\bibitem[Yue et al.(2013b)]{2013MNRAS.433..1556Y} Yue, B., Ferrara, A., Salvaterra, R., Xu, Y., et al.\ 2013, \mnras, 433, 1556
\bibitem[Zemcov et al.(2013)]{2013ApJS..207...31Z} Zemcov, M., Arai, T., Battle, J., et al.\ 2013, \apjs, 207, 31
\bibitem[Zemcov et al.(2014)]{2014Sci...346..732Z} Zemcov, M., Smidt, J., Arai, T., et al.\ 2014, {\it Science}, 346, 732
\bibitem[Zubko et al.(2004)]{2004ApJS..152..211Z} Zubko, V., Dwek, E., \& Arendt, R. G.\ 2004, \apjs, 152, 211

\end{thebibliography}
\end{document}